\documentclass[final,5p,times,twocolumns]{elsarticle}

\usepackage{graphicx}

\usepackage{amssymb}

\usepackage[modulo]{lineno}

\usepackage{amsmath,amsthm}
\usepackage[colorlinks=true, allcolors=blue]{hyperref}
\usepackage{multirow}
\usepackage{caption}
\usepackage{booktabs}
\usepackage{algorithm}
\usepackage{algorithmicx}
\usepackage{algpseudocode}
\usepackage{newtxtext,newtxmath}
\usepackage[usenames,dvipsnames]{xcolor}

\DeclareSymbolFont{letters}{OML}{cmm}{m}{it}
\DeclareMathAlphabet{\mathcal}{OMS}{cmsy}{m}{n}

\biboptions{sort&compress}

\begin{document}

\begin{frontmatter}

\title{Multi-level Delumping Strategy for Thermal Enhanced Oil Recovery \\ Simulations at Low Pressure}

\author[label1]{Matthias A. Cremon\corref{cor1}}
\address[label1]{Department of Energy Resources Engineering, Stanford University}

\cortext[cor1]{Corresponding author.}

\ead{mcremon@stanford.edu}

\author[label1]{Margot G. Gerritsen}
\ead{gerritsen@stanford.edu}

\begin{abstract}

We present a multi-level delumping method suitable for thermal enhanced oil recovery processes. At low pressures, the temperature variable is the most critical factor impacting the displacement process through viscosity reduction and evaporation/condensation effects. Hydrocarbon components are vaporized under high temperatures, move downstream in the gas phase and condense back to the liquid phase. That process is governed by the $K$-values of the components, evaporating out of the liquid phase sequentially with increasing temperatures. To reduce the computational cost, it is standard practice to reduce the number of (pseudo-)components used in thermal reservoir simulation. Depending on the number and type of hydrocarbon pseudo-components retained in the simulations, we may not be able to capture the correct displacement due to large errors in the lumped phase behavior (flash) computations. We address that problem through a multi-level method: we use data obtained from a short simulation using the most detailed fluid description available, and leverage that information to guide a delumping process. We use temperature as a proxy variable for composition, and select reference temperatures. We extract the corresponding reference compositions from the detailed run and use them to extend the lumped pseudo-components to an approximate detailed composition. We compute the phase mole fractions as well as the gas compressibility factor. We test our method using six heavy oil samples, and under two different recovery processes: hot nitrogen injection and in-situ combustion (air injection and exothermic oxidation reactions). The average error on the liquid mole fraction is reduced by 4-12 times (depending on the oil samples) compared to the flash using pseudo-components, and the maximum error by 6-48 times. We illustrate that the method is amenable to manually adding more information about the physics of some oil samples. We also discuss how to efficiently pick the reference temperatures. For uniformly sampled temperatures (between a minimum and maximum temperature), we conduct a sensitivity study which led us to use six temperatures. We ran both local (Pattern Search, PS) and global (Particle Swarm Optimization, PSO) gradient-free optimization methods. PS is able to find the closest local minimum to the uniform set, giving a limited improvement of 6.5\%. The known increased cost for PSO is worth the investment in at least one of the cases we considered, leading to a 67\% improvement.


\end{abstract}


\end{frontmatter}


\newpage

\section{Introduction}
\label{sec::intro}

Heavy and extra-heavy crude oils can be characterized using detailed fluid descriptions comprising tens of hydrocarbon components. Thermal reservoir simulation is computationally expensive and it is intractable to use so many components with thousands of grid cells. As a result, it is standard practice to use a lumping procedure to reduce the number of components, usually to around 3-7 hydrocarbon pseudo-components. Reactive simulations capable of modeling combustion processes are even more costly due to the increased coupling and non-linearity of the problem. They also require more non-hydrocarbon components (typically nitrogen, oxygen, carbon monoxide and carbon dioxide, as well as water), leading to a larger linear system.

Lumping has been extensively used in the petroleum community, and thermodynamically consistent properties can be obtained for pseudo-components using pure component mixing rules  \citep{Kay36,Montel84,Nowley91,Leibovici93,Jessen07,Rastegar09,Pedersen14}. There is a loss of information from lumping and the accuracy of the phase behavior computations will decrease. In some cases, we can rigorously get back information about the detailed composition by using a delumping procedure. The delumping literature typically assumes that the displacement process in the reservoir is not affected by the number of components \citep{Leibovici96,Nichita06a,Nichita06b,Nichita07,DeCastro11}. In other words, the number and nature of pseudo-components used for the lumped case is enough to represent the phase behavior for the entire displacement process. Unfortunately, for thermal, compositional, reactive simulations, the minimum number of pseudo-components needed to capture the displacement can be above ten due to the wider range of temperatures experienced by the mixtures, and their different compositions \citep{Cremon20d}. The resulting simulation time using around 15 total components is likely to be on the order of days for a grid with only thousands of cells with current reservoir simulators \citep{stars,eclipse,adgprs}.

In this work, we take a different approach and attempt to delump the global molar fractions using information from a reference run. Using the proposed method requires running part of a detailed (meaning using the largest number of components available) simulation. For laboratory scale In-Situ Combustion (ISC), pressure is low and roughly constant, and the displacement is largely governed by the evaporation and condensation of hydrocarbon components. Both of those facts lead us to use the temperature as a proxy variable for composition. Doing so allows us to improve the accuracy of the phase behavior computations by delumping the pseudo-components according to a reference temperature-dependent composition. Specifically, we want to compute the phase molar fractions, $\textrm{L}$, $\textrm{V}$, $\textrm{W}$, and the gas phase compressibility $Z_v$ with higher accuracy, so that the displacement process could be captured accurately, but for a fraction of the detailed simulation cost. The reduced case makes this method attractive for uncertainty quantification, calibration of parameters or sensitivity analyses -- all of which require a large number of simulations.

\section{Thermal Enhanced Oil Recovery Processes}

Recovering heavy and extra-heavy oils is typically done using tertiary recovery processes, such as thermal Enhanced Oil Recovery (EOR) methods \citep{Prats82}. Thermal EOR involves the injection of heat into a reservoir, usually delivered through a hot fluid (water or steam), or the generation of heat in-situ. In-situ Combustion (ISC) consists of the injection of air in the reservoir combined with local heating. Under such conditions, part of the oil will be oxidized, releasing large amounts of heat. That heat will lower the viscosity of the oil, allowing it to be displaced by the injected gases.

All of the thermal EOR processes can be described using the same set of partial differential equations (PDEs), given in the following subsections.

\subsection{Conservation Equations}

The mass is conserved for each component across all four phases (oil, water, gas and solid). First, we consider $n_c$ fluid components in $n_p$ fluid phases in the general case, leading to
\begin{align} \label{eq::mass}
\begin{split}
 \dfrac{\partial}{\partial t}\left(\phi\sum\limits_{p\,=\,1}^{n_p}x_{cp}\rho_pS_p\right) + \nabla \cdot \left(\sum\limits_{p\,=\,1}^{n_p}x_{cp}\rho_p\textbf{u}_p\right) + q_c^\textrm{w} + q_c^\textrm{r} = 0,
\end{split}
\end{align}
for $c = 1,\dots,n_c,$ and where $p$ and $c$ are the phase and component indices, $\phi$ is the porosity, $x_{cp}$ is the mole fraction of component $c$ in phase $p$, $\rho_p$, $S_p$ and $\textbf{u}_p$ are the molar density, saturation and velocity of phase $p$, $q_c^\textrm{w}$ is the source term from wells and $q_c^\textrm{r}$ the source term from reactions. We compute $\textbf{u}_p$ using the standard multi-phase extension of Darcy's law \citep{Darcy56,Muskat36}. We also consider $n_s$ solid components obeying
\begin{equation} \label{eq::mass_solid}
 \dfrac{\partial}{\partial t}\left(\phi c_s\right) + q_s^\textrm{r} = 0,
\end{equation}
for $s = 1,\dots,n_s,$ where $c_s$ is the molar concentration and $q_s^\textrm{r}$ the source term from reactions.

In the thermal formulation, we also need to conserve energy according to
\begin{align} \label{eq::energy}
\begin{split}
\dfrac{\partial}{\partial t}\left(\phi\sum\limits_{p\,=\,1}^{n_p}U_{p}\rho_pS_p+(1-\phi)\tilde U_R\right) + \nabla \cdot \left(\sum\limits_{p\,=\,1}^{n_p}H_{p}\rho_p\textbf{u}_p\right) \\ - \nabla \cdot \left(\kappa\nabla \textrm{T}\right) + q^\textrm{w} + q^\textrm{r} + q^\textrm{hl} + q^\textrm{hr} = 0,
\end{split}
\end{align}
where $p$ and $c$ are the phase and component indices, T is the temperature, $\tilde U_R$ is the rock volumetric internal energy, $\kappa$ is the thermal conductivity, $U_{p}$ and $H_{p}$ are the internal energy, enthalpy of phase $p$, $q^\textrm{w}$, $q^\textrm{r}$, $q^\textrm{hl}$ and $q^\textrm{hr}$ are the source terms from wells, reactions, heat losses and heater, respectively.

In general, most of the properties appearing in the conservation equations are functions of pressure $(\textrm{P})$, temperature $(\textrm{T})$ and global compositions ($z_i$), through phase behavior calculations. The phase molar fractions and compressibility factors do not appear in Eqs.~\eqref{eq::mass}--\eqref{eq::energy}, but they are at the root of density, saturation and enthalpy calculations. An inaccurate phase behavior routine will prevent us from capturing the correct physical displacement.

\subsection{Phase Behavior Model}

We use a free-water (FW) flash for the phase behavior calculations, initially presented in \citet{Lapene10}. For heavy oil recovery using thermal methods, the solubility of oil in water is negligible. Therefore, we use the following equations to partition the overall molar fractions
\begin{align}
z_i & = \,\, x_i\textrm{L} + \ y_i\textrm{V}, \hspace{1cm} i = 1,\dots,n_c,\ i \neq w  \,, \label{eq::FW1} \\
z_w & = x_w\textrm{L} + y_w\textrm{V} + \textrm{W}, \label{eq::FW2}
\end{align}
with $\textrm{L}$, $\textrm{V}$ and $\textrm{W}$ the oil, vapor and water phase molar fractions, subscript $_w$ the water component index, $z_i$ the overall mole fraction of component $i$ and , $x_i$, $y_i$ and $w_i$ the liquid, vapor and water mole fractions of component $i$, respectively. These equations state that the hydrocarbon components cannot be dissolved in the water phase and that the water component can be present in all phases, respectively. \citet{Lapene10} tested the method on several oils including one used in this work, and reported excellent agreement with a full three-phase flash \citep{Petitfrere14}.

\subsection{Composition Variability in Time and Space}

\begin{figure*}[t!]
    \centering
    \includegraphics[width=\textwidth]{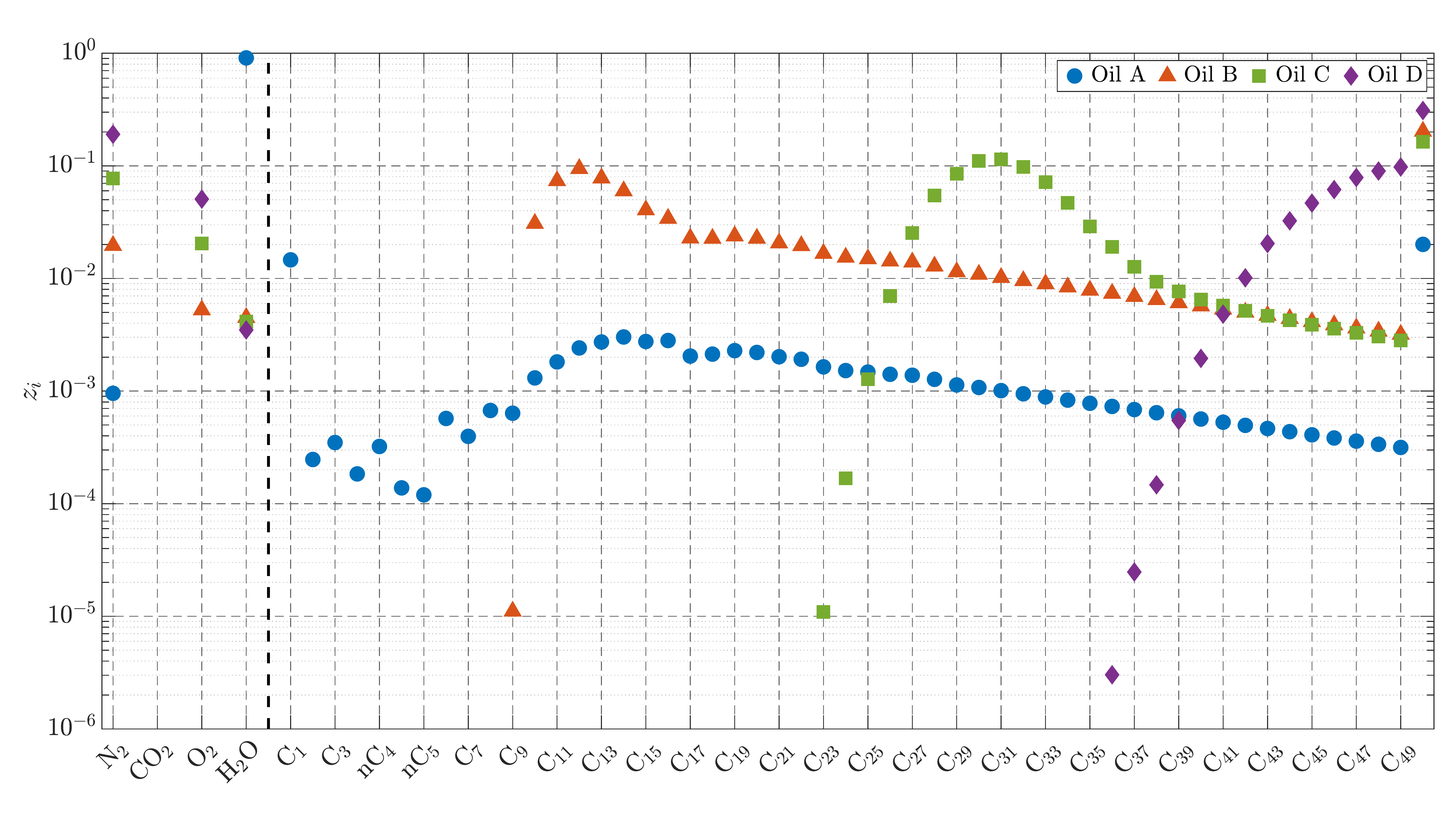}
    \caption{Global molar composition ($z_i$) for different oil compositions: oil A (blue dots), oil B (orange triangle) oil C (green squares) and oil D (purple diamonds). All compositions come from the same block of a hot air injection simulation, and correspond to different timesteps.}
    \label{fig::oilSamples}
\end{figure*}

Lumped pseudo-components are typically constructed in a way that approximates the phase envelope of the initial oil as well as possible, based on experimental data. During thermal recovery, components get vaporized and the oil composition will vary both spatially and over time. In those cases, using a low number of pseudo-components is unlikely to be able to approximate the phase behavior of a complex heavy oil accurately. Figure~\ref{fig::oilSamples} shows four compositions we observe in one block during a hot air injection simulation. Oil A is the initial extra-heavy oil, described at length in \citet{Lapene10b}, where we added 90\% water content by mole. That oil has light, medium and heavy components, as well as a significant portion of its heavy fraction $\textrm{C}_{50+}$ (around 2\% by mole overall, 22\% of the hydrocarbon content). Our initial conditions are $50^\circ\textrm{C}$ and $8$ bars. Oil B is from an intermediate timestep; the conditions are $143^\circ\textrm{C}$ and $9$ bars. We see that the light components have all been vaporized, and intermediate ones (roughly $\textrm{C}_{10}$ to $\textrm{C}_{17}$) are banking and make up, along with the plus fraction, most of the hydrocarbon content. Oil C is from a late time, corresponding to $377^\circ\textrm{C}$ and $9.4$ bars. Every hydrocarbon lighter than $\textrm{C}_{23}$ has been evaporated. Finally, Oil D is the latest time reported here, subjected to $445^\circ\textrm{C}$ and $9.7$ bars. It shows no hydrocarbon components below $\textrm{C}_{36}$.

Those four compositions show very different phase behaviors. Oils B, C and D give only two phases, with water present only in the vapor phase. The later time oils have more nitrogen, leading to a growing vapor phase fraction. We define the error on the liquid mole fraction $(\textrm{L})$ as the 1-norm of the difference with the reference, detailed composition results:
\begin{equation}
    \epsilon = |\textrm{L} - \textrm{L}_\textrm{ref}|
\end{equation}

The detailed 52-component description would be lumped into two to eight hydrocarbon pseudo-components for reservoir simulations \citep{Lapene10b}.
\begin{table}[t]
    \centering
    \caption{Liquid (L), Vapor (V) and Water (W) molar fractions, and error on the liquid mole fraction ($\epsilon$) using 4 lumped pseudo-components for our four oil compositions.}
    {
    \begin{tabular}{lcccc}
    \toprule
        Composition & L & V & W & $\epsilon$\\
    \midrule
        Oil A & 0.09063 & 0.00288 & 0.90649 & 0.00014 \\
        Oil B & 0.99296 & 0.00704 & 0.00000 & 0.00026 \\
        Oil C & 0.92195 & 0.07806 & 0.00000 & 0.00203 \\
        Oil D & 0.77648 & 0.22352 & 0.00000 & 0.02232 \\
    \bottomrule
    \end{tabular}
    }
    \label{tab::samples}
\end{table}
Table~\ref{tab::samples} shows the flash results for all four oil samples: the mole fractions of each phase, as well as the error on the liquid fraction when using a four pseudo-components lumping $(\textrm{C}_1$, $\textrm{C}_{2-16}$, $\textrm{C}_{17-49}$, $\textrm{C}_{50+})$. Four pseudo-components is a common breakdown, since it allows to label them as gas (mostly methane), light oil, heavy oil and the plus fraction. There are many ways of assigning components to those pseudo-components \citep{Leibovici93,Jessen07,Rastegar09}. In this work, we use the method described in \citet{Leibovici93}. As we expected, the performance of the lumped flash deteriorates with later timesteps and higher temperatures. The lumping hides the repartition of the heaviest components into one $(\textrm{C}_{17-49})$ pseudo-component, which does not allow to capture the correct phase properties.

\begin{figure*}[t!]
    \centering
    \includegraphics[width=\textwidth]{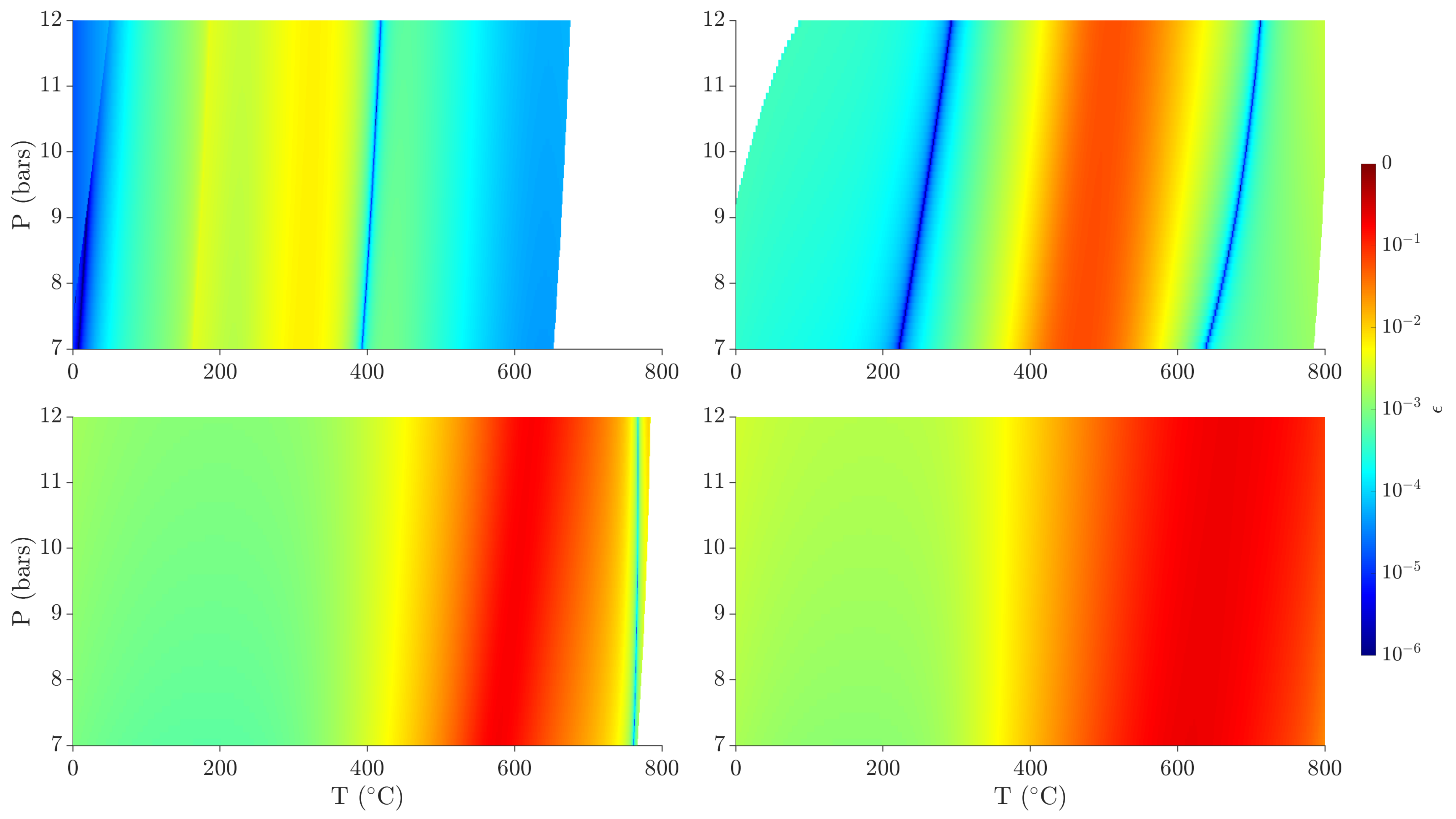}
    \caption{1-norm error on the liquid mole fraction (L) using four pseudo-components $(\textrm{C}_1$, $\textrm{C}_{2-16}$, $\textrm{C}_{17-49}$, $\textrm{C}_{50+})$ for oil A (top left), B (top right), C (bottom left) and D (bottom right).}
    \label{fig::oil1enveloppe}
\end{figure*}

We compute the errors made on the liquid fraction for several pressures and temperatures and plot the results on Figure~\ref{fig::oil1enveloppe}. The conditions are consistent with what we observe in the simulations. As we can see, the four components lumping performs relatively well on the initial oil, with a maximum error of $\mathcal{O}\left(10^{-2}\right)$. The accuracy drops sharply with Oil B, C and D, all showing large regions of $\mathcal{O}\left(10^{-1}\right)$ error, and even $\mathcal{O}\left(1\right)$ for the maxima. Errors on the phase mole fractions will lead to inaccurate calculations of phase properties, such as phase densities and saturations, which will in turn strongly impact the non-linear displacement process.

Running cases with the detailed composition, using in our case 52 hydrocarbon components, is not tractable in the context of calibration, uncertainty quantification or experimental design. Delumping the composition at the flash level allows us to retain only a small number of pseudo-components for the displacement, keeping the Jacobian size and the computational cost to a minimum. Most of the delumping work \citep{Leibovici96,Nichita06a,Nichita06b,Nichita07,DeCastro11} has been done under the assumption that the detailed components behave like tracers with respect to the lumped components. For the cases we are interested in, depending on the oil samples, we would need a total number of (pseudo-)components in the 15--20 range to satisfy that assumption. We will therefore pursue another method, which uses knowledge about the physics and a single detailed simulation to guide the lumped phase behavior calculations. The idea is similar to a parametrization of the compositional space -- by a projection to the one-dimensional temperature space. Similar ideas have been successfully applied to compositional simulations \citep{Voskov09b,Iranshahr10,Ganapathy18,Chen19} or $K$-value calculations \citep{Rannou10,Rannou13}.

\section{Multi-level Delumping Strategy}

The objective of our delumping procedure is to estimate the mole fractions of the detailed components based on the lumped pseudo-components and a reference detailed composition. We first describe the delumping methodology, and then expand on the way we select the required reference compositions.

\subsection{Delumping Procedure}

We consider the lumped composition $\tilde z_j$, where $j$ is the pseudo-component index in the lumped composition. We know that each lumped pseudo-components corresponds to several detailed components. We also need a reference detailed composition, $z^\textrm{r}_{i}$, where $i$ is the component index in the detailed composition. We denote $\tilde z^\textrm{r}_{j}$ the lumped reference composition, and compute it as
\begin{equation}
    \tilde z^\textrm{r}_{j} = \sum\limits_{p\,\in\,\mathcal{I}_j} z_p^\textrm{r}, \quad j = 1,\dots,n_{pc},
\end{equation}
where $\mathcal{I}_j$ is the ensemble of detailed component indices $i$ corresponding to the $j^{\textrm{th}}$ lumped pseudo-component and $n_{pc}$ the number of pseudo-components.

We can then expand the lumped composition into the de-lumped composition, using
\begin{equation}
    z_i^\textrm{dl} = \tilde z_{j}\dfrac{z_i^\textrm{r}}{\tilde z^\textrm{r}_{j}}, \quad i = 1,\dots,n_c,
\end{equation}
where $z^\textrm{dl}$ denotes the delumped composition and $n_c$ is the number of detailed components. It is important to note that this method is mass conservative by construction, since we only map the pseudo-components onto the detailed components using the reference detailed composition. 

If that reference composition is a good representation of the detailed composition, we will be able to add back some of the details lost in the lumping and enhance the quality of the flash results. It should also be noted that the method can lead to worse results than the lumped flash if the reference composition is picked poorly.

\begin{figure*}[htb!]
    \centering
    \includegraphics[width=\textwidth]{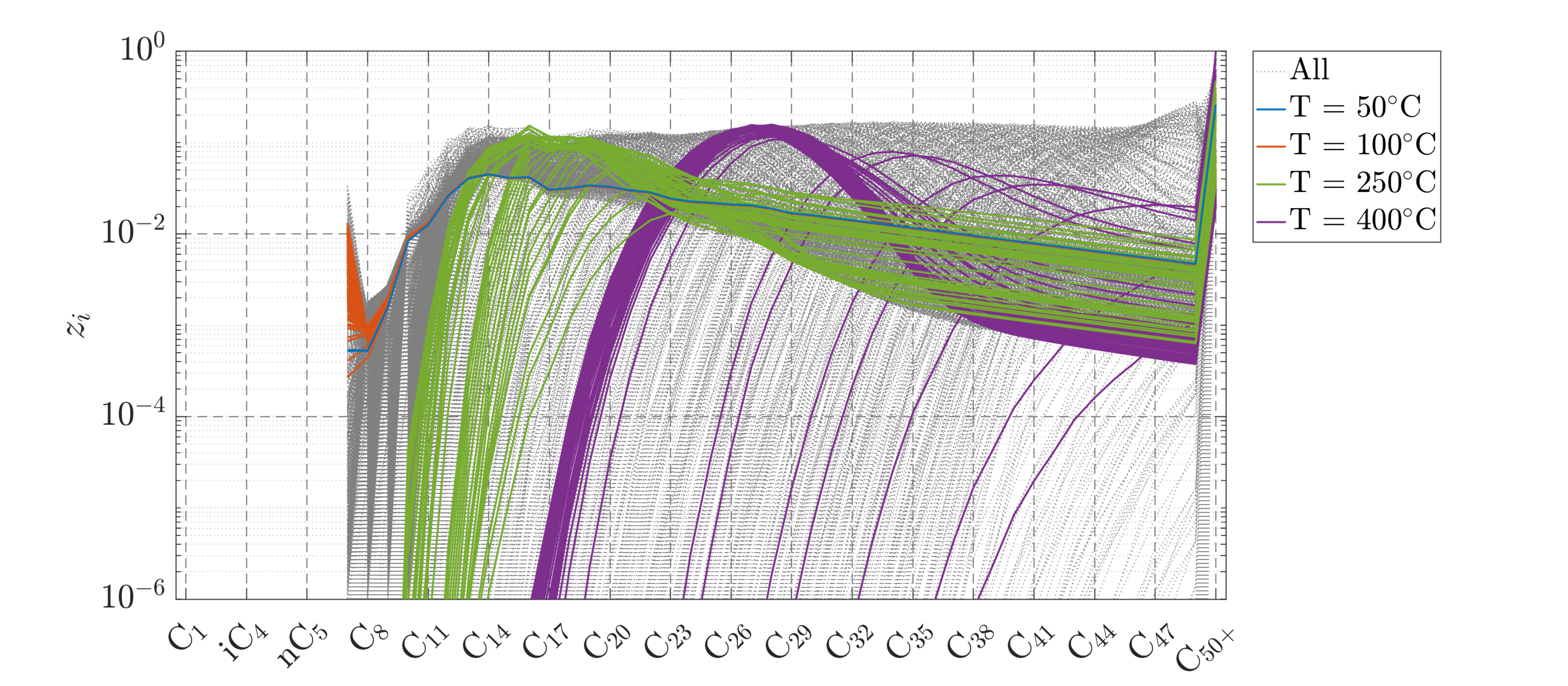}
    \caption{Hydrocarbon compositions observed in the simulation. The dashed grey lines show all compositions, and the plain lines are temperatures $\pm2^\circ$C of 50$^\circ$C (blue), 100$^\circ$C (orange), 250$^\circ$C (green) and 400$^\circ$C (purple).}
    \label{fig::allCompositionsTemp}
\end{figure*}

In the next section, we detail how and why we select the reference compositions $z^\textrm{r}$.

\subsection{Multi-level Strategy}


Our multi-level method relies on some of the physical properties of thermal recovery processes. When heat is injected or generated in an oil reservoir, multiple fronts will be moving downstream. Depending on the case and conditions, we can observe a steam front, a water evaporation front, an oil evaporation front, a reaction front and several $K$-value fronts. 

For the cases we are interested in, namely hot gas injection and in-situ combustion, the displacement is mostly governed by the evaporation and condensation of components through phase behavior, and the viscosity reduction. Both of those are strong functions of temperature. Moreover, at laboratory conditions, the pressure is low (below 10 bars) and virtually constant. The combination of those reasons lead us to use temperature as the proxy variable for composition. In other words, we are relying on the fact that for a given oil at a given temperature occurring during a thermal recovery process, the oil composition is likely to look the same regardless of time and space. In Figure~\ref{fig::allCompositionsTemp}, the gray dashed lines show all the compositions we experience over an ISC simulation (every block, every time step). The colored lines shows the compositions of blocks within 2$^\circ$C of 400$^\circ$C (purple), 250$^\circ$C (green), 100$^\circ$C (orange) and 50$^\circ$C (blue). We observe clear clustering, with lower temperatures showing virtually a single curve. Higher temperatures show a slightly larger spread, but still small compared to the range of compositions experienced over the course of the simulation. Due to the identified relationship between temperature and composition, we will select a few representative temperatures and extract the corresponding compositions from the detailed simulation.

\begin{figure}[htb!]
    \centering
    \includegraphics[width=.48\textwidth]{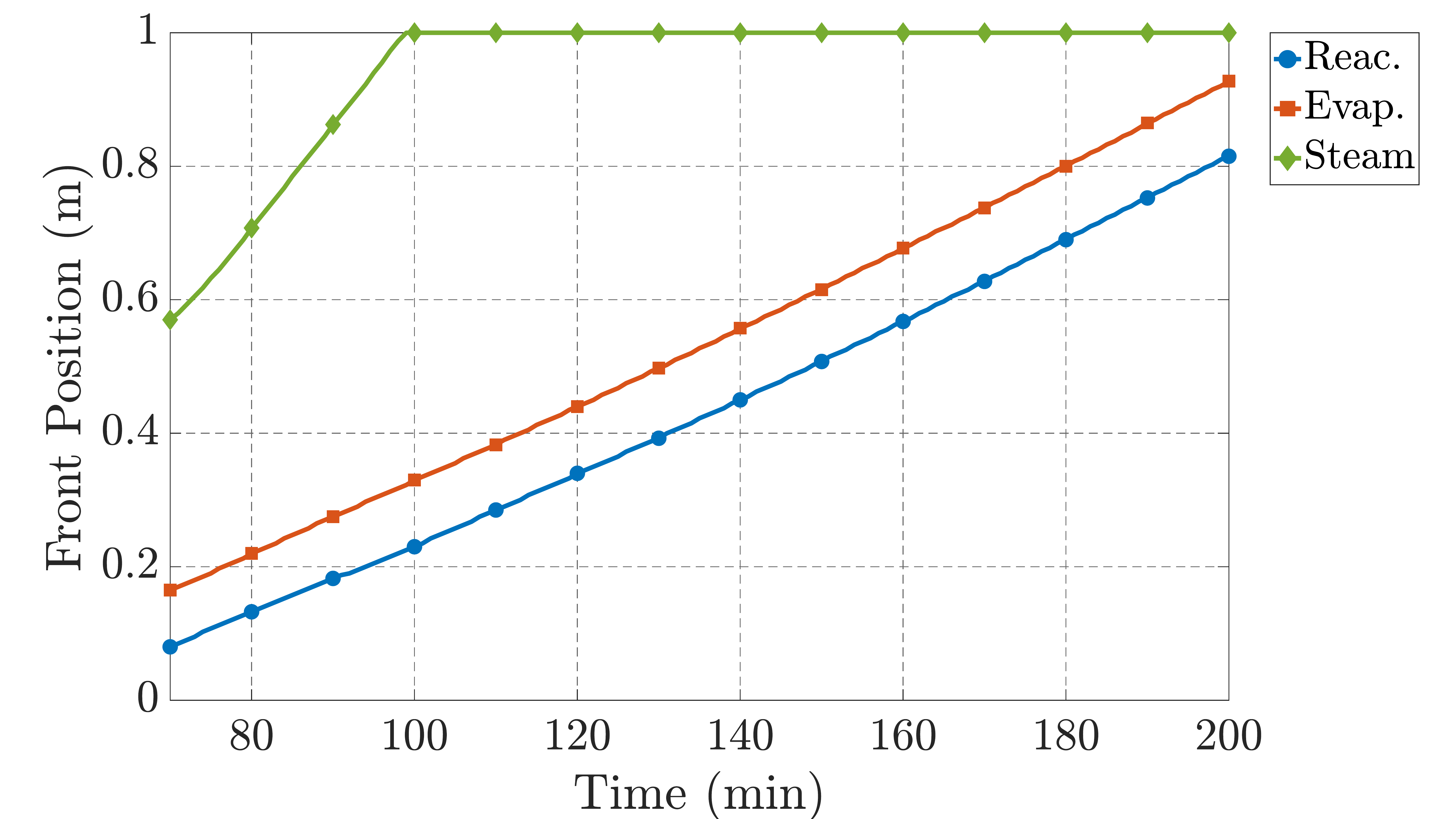}
    \caption{Positions over time for the reaction (blue circles), water evaporation (orange squares) and steam (green diamonds) fronts. All of them show a linear profile, indicating a constant propagation speed.}
    \label{fig::fronts}
\end{figure}

This strategy is further justified by the nature of the displacement process. The temperature fronts moving downstream exhibit a self-similar behavior. Although due to the complexity of the coupled non-linear equations, we cannot conduct a fractional-flow-based study of the hyperbolic problem \citep{Ganapathy18,Chen19}, our numerical results show that multiple temperature shocks are moving downstream, namely a steam front, a water-evaporation front, and a reaction (or maximum temperature) front. Figure~\ref{fig::fronts} shows the position of the fronts over time for a combustion simulation with Dead Zuata oil. We see that after ignition (which occurs around 60 minutes), all fronts show a linear trend indicating that they move at a constant speed and the solution is near self-similar.

\begin{figure}[tb!]
    \centering
    \includegraphics[width=.48\textwidth]{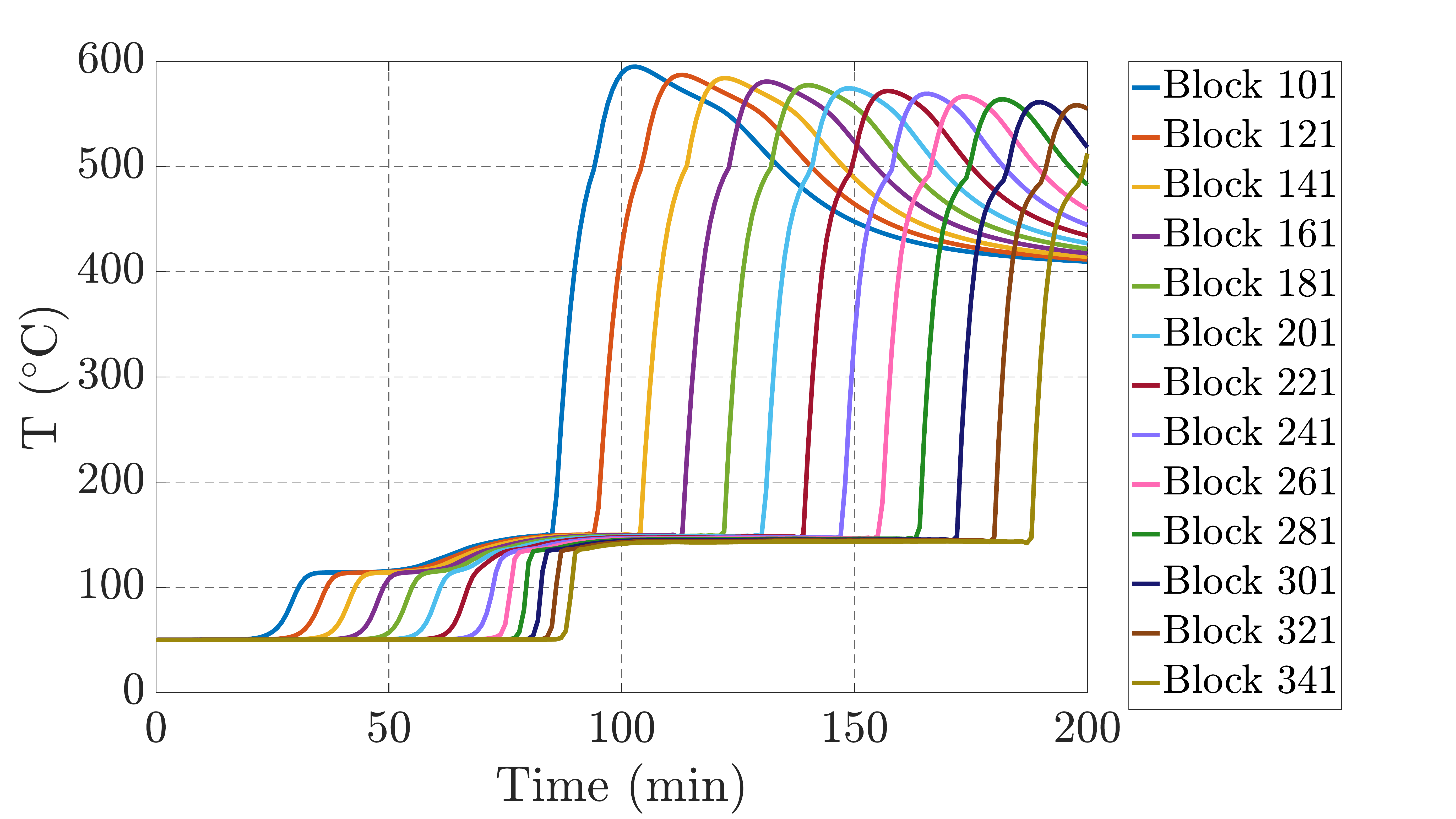}
    \caption{Temperature profiles over time, for several blocks of an air injection simulation using Dead Zuata.}
    \label{fig::tempProfiles}
\end{figure}

Figure~\ref{fig::tempProfiles} shows the temperature profiles over time in several blocks for an ISC case and confirms the previous statements. We observe that the blocks experience a very similar temperature history, with a delay corresponding to the speed at which the temperature shocks are traveling. For that reason, we can select a reference block as early as possible in the domain (provided that it experiences the relevant temperature history) and only use information from that block, thus limiting the storage requirements as well as the runtime needed for the detailed simulation.

\subsection{Reference Temperature Selection}

We now discuss the way we select the reference temperatures used in the delumping process. In many delumping methods, the initial composition is used to map the detailed and lumped compositions. We will consider that case, which amounts to using only one temperature in our method, and set it to the initial temperature (50$^\circ$C). We call that method initial in the remainder of this work. We also compared with the regular lumped flash.

Then, we need to select how many temperatures to use in the multi-level scheme, and which ones. Our base case is to use six temperatures, uniformly picked between the initial condition (50$^\circ$C) and a maximum temperature of 400$^\circ$C for nitrogen injection cases, and 450$^\circ$C for air injection cases. Above those temperatures, in most cases only the very heavy components are left, usually only the plus fraction -- which is not a lumped component, so no additional lumping information is available. We need a larger temperature for air injection cases, because the temperature increase occurs faster due to reactions, so more components are left in the mixture at a given temperature. In most of our test cases, the uniform sampling method performs well. We also optimized the selection of the temperatures, and present those results in Section~\ref{sec::optim}.

\begin{figure*}[t!]
    \centering
    \includegraphics[width=\textwidth]{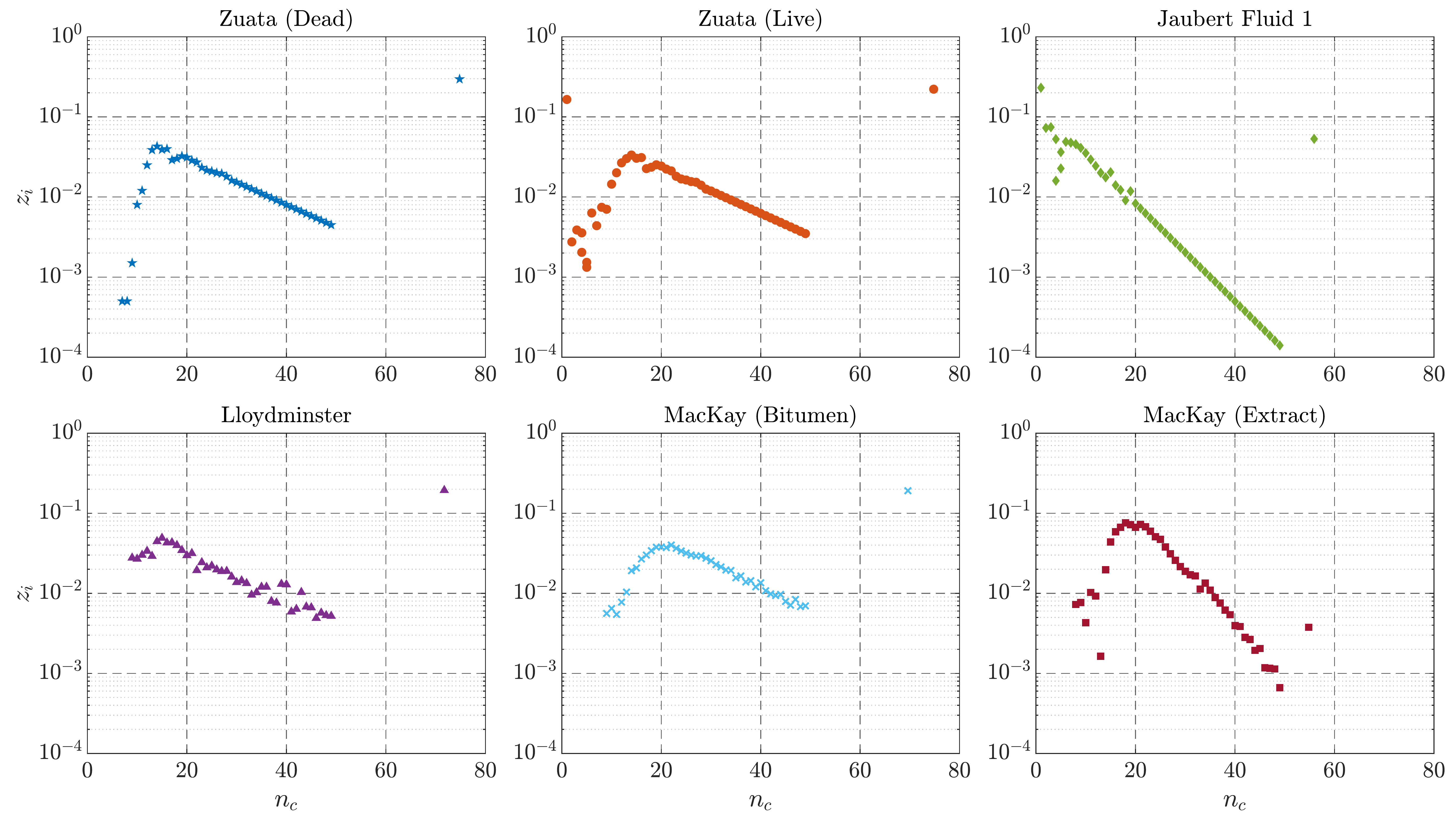}
    \caption{Molar fractions of the detailed hydrocarbon components for all oil samples: Dead Zuata (top left), Live Zuata (top center), Jaubert Fluid 1 (top right), Lloydminster (bottom left), MacKay Bitumen (bottom center) and MacKay Extract (bottom right).}
    \label{fig::compositions}
\end{figure*}

\section{Numerical Results}
\label{sec::results}

We tested our method on six different heavy oil samples, which we briefly describe in the next section.

\subsection{Oil Samples}

The first two oils are from the Zuata field, in the Orinoco Belt region of Venezuela. They are both described in \citet{Lapene10b}. We consider the laboratory condition oil, denoted Dead, and a recombined oil designed to approximate the oil composition at reservoir conditions (around 48$^\circ$C and 40 bars), denoted Live. They were both characterized until C$_{30+}$, and then extended to the C$_{50+}$ fraction. The Dead oil shows no light components, $\textrm{C}_7$ being the first hydrocarbon present, but the Live oil was blended with methane (16\% by mole) and other light components. The Zuata oils are the heaviest samples, with a density of 8.5$^\circ\!$API.

Our third sample is described in \citet{Jaubert02a} and referred to as Fluid 1 in the paper. The description is given up to C$_{20+}$, and we extended it to C$_{50+}$. Although the lightest oil we consider at around 35$^\circ\!$API, it contains about 6\% by mole of C$_{50+}$ and shows a virtually log-linear repartition of other components. According to \citet{Jaubert02a}, that oil was produced by miscible CO$_2$ injection.

The fourth oil is a bitumen from Lloydminster, Saskatchewan (Canada), studied in \citet{Li13}. It is similar to Dead Zuata, with no light components and a large C$_{50+}$ content. The composition was obtained by simulated distillation (no extension) and is given up to C$_{60+}$, which we relump to C$_{50+}$ for convenience and consistency with other samples. Its density is 10.0$^\circ\!$API.

Finally, the fith and sixth samples are described in \citet{Nourozieh13} and come from MacKay, Alberta (Canada). We selected two samples from that work, obtained by mixing a bitumen with an ethane solution; the mixture forms two liquid phases. The heaviest is denoted Bitumen, and the lightest Extract. The bitumen is once again similar to the Zuata and Lloydminster samples and slightly lighter at 12.9$^\circ\!$API. The extract however is quite interesting. It can be characterized as heavy with its density of 18.0$^\circ\!$API, but the composition shows an inverted-U shape with few light and heavy components, but 50\% of the composition in the $\textrm{C}_{15}$-$\textrm{C}_{35}$ regions.

\begin{table}[b!]
    \centering
    \caption{Summary of the oil samples properties.}
    {
    \begin{tabular}{lrrr}
    \toprule
        Oil & Country & Category & API  \\
    \midrule
       Dead Zuata & Venezuela & Extra-heavy & 8.5$^\circ$  \\
       Live Zuata & Venezuela & Extra-heavy & 8.5$^\circ$  \\
       Jaubert & --- & Medium & 35.0$^\circ$  \\
       Lloydminster & Canada & Bitumen & 10.0$^\circ$  \\
       MacKay Bitumen & Canada & Bitumen & 12.9$^\circ$  \\
       MacKay Extract & Canada & Heavy & 19.0$^\circ$  \\
    \bottomrule
    \end{tabular}
    }
    \label{tab::oils}
\end{table}

A brief summary of the oil samples properties is given in Table~\ref{tab::oils}, and all of the compositions are shown in Figure~\ref{fig::compositions}. We mix the oils with 90\% water by mole and 1\% nitrogen -- with the exception of the Jaubert Fluid 1 sample, where we add 90.99\% water and 0.01\% nitrogen so that the oil and water saturations are higher. To test our delumping algorithm, we will use the output data from our detailed simulation cases. For each timestep and each block, we extract the pressure, temperature and overall composition. We then flash that composition, to obtain the reference molar fractions and compressibility factors (using a standard Peng-Robinson \citep{Peng76} equation of state). We also compute the pseudo-components compositions, and flash the corresponding lumped composition. Finally, we delump the lumped composition using both the initial detailed composition only, and our multi-level method. In this work, our 1D simulation cases have 400 blocks and we run 201 timesteps (including the initialization). We consider $201\times 400 = 80,\!400$ test samples for each oil, unless otherwise specified. We use non-zero binary interaction coefficients (BICs) for some species, as shown in Table~\ref{tab::BICs}.

\begin{table}[htb]
    \centering
    \caption{Non-zero binary interaction coefficients (BICs).}    
    \begin{tabular}{lllll}
    \toprule
        Component $j$ & $k_{\textrm{N}_2-j}$ & $k_{\textrm{CO}_2-j}$ & $k_{\textrm{H}_2\textrm{O}-j}$ & $k_{\textrm{C}_1-j}$ \\
     \midrule
        $\textrm{N}_2$ & $-$ & 0.1000 & 0.4778 & 0.1000 \\
        $\textrm{CO}_2$ & 0.1000 & $-$ & 0.1896 & 0.1200 \\
        $\textrm{H}_2\textrm{O}$ & 0.4778 & 0.1896 & $-$ & 0.4850 \\
        $\textrm{C}_1$ & 0.1000 & 0.1200 & 0.4850 & $-$ \\
        $\textrm{C}_2$ & 0.1000 & 0.1200 & 0.5000 & 0 \\
        $\ \ \vdots$ & $\ \ \vdots$ & $\ \ \vdots$ & $\ \ \vdots$ & $\vdots$  \\
        $\textrm{C}_{10}$ & 0.1000 & 0.1200 & 0.5000 & 0 \\
        $\textrm{C}_{11}$ & 0.1000 & 0.1200 & 0.5000 & 0.0700 \\
        $\ \ \vdots$ & $\ \ \vdots$ & $\ \ \vdots$ & $\ \ \vdots$ & $\ \ \vdots$  \\
        $\textrm{C}_{50+}$ & 0.1000 & 0.1200 & 0.5000 & 0.0700 \\
    \bottomrule
    \end{tabular}
    \label{tab::BICs}
\end{table}

\subsection{Test Case 1: Hot Nitrogen Injection}
\label{subsec::nit}

Our first tests use data from one dimensional hot nitrogen injection cases. Since there is no oxygen, all reactions are inactive. This case represents the initial stage of a combustion case, but for the sake of testing we do not switch to air injection so that we can see the propagation of the different fronts. We inject hot nitrogen (600$^\circ$C) into our mixture of oil, water and nitrogen. The initial conditions are 50$^\circ$C and 8 bars, and the injection conditions 3L/min. The main parameters of the simulations are summarized in Table~\ref{tab::simparams}.

\begin{table}[b!]
    \centering
    \caption{Parameters for hot nitrogen injection simulations.}
    {
    \begin{tabular}{llrl}
    \toprule
    Property & Symbol & Value & Unit \\
    \midrule
    Domain Size & d & 1 & m \\
    Porosity & $\phi$ & 0.36 & -- \\
    Permeability & $k$ & 10 & D \\
    Injection Rate & $q$ & 3 & L/min \\
    Injection Temperature & T$_\textrm{inj}$ & 600 & $^\circ$C \\
    Initial Temperature & T$_\textrm{init}$ & 50 & $^\circ$C \\
    Initial Pressure & P$_\textrm{init}$ & 8 & bar \\
    Number of Blocks & $n_\textrm{b}$ & 400 & -- \\
    Total Time & $t$ & 200 & min \\
    \bottomrule
    \end{tabular}
    }
    \label{tab::simparams}
\end{table}


\begin{figure*}[t]
    \centering
    \includegraphics[width=\textwidth]{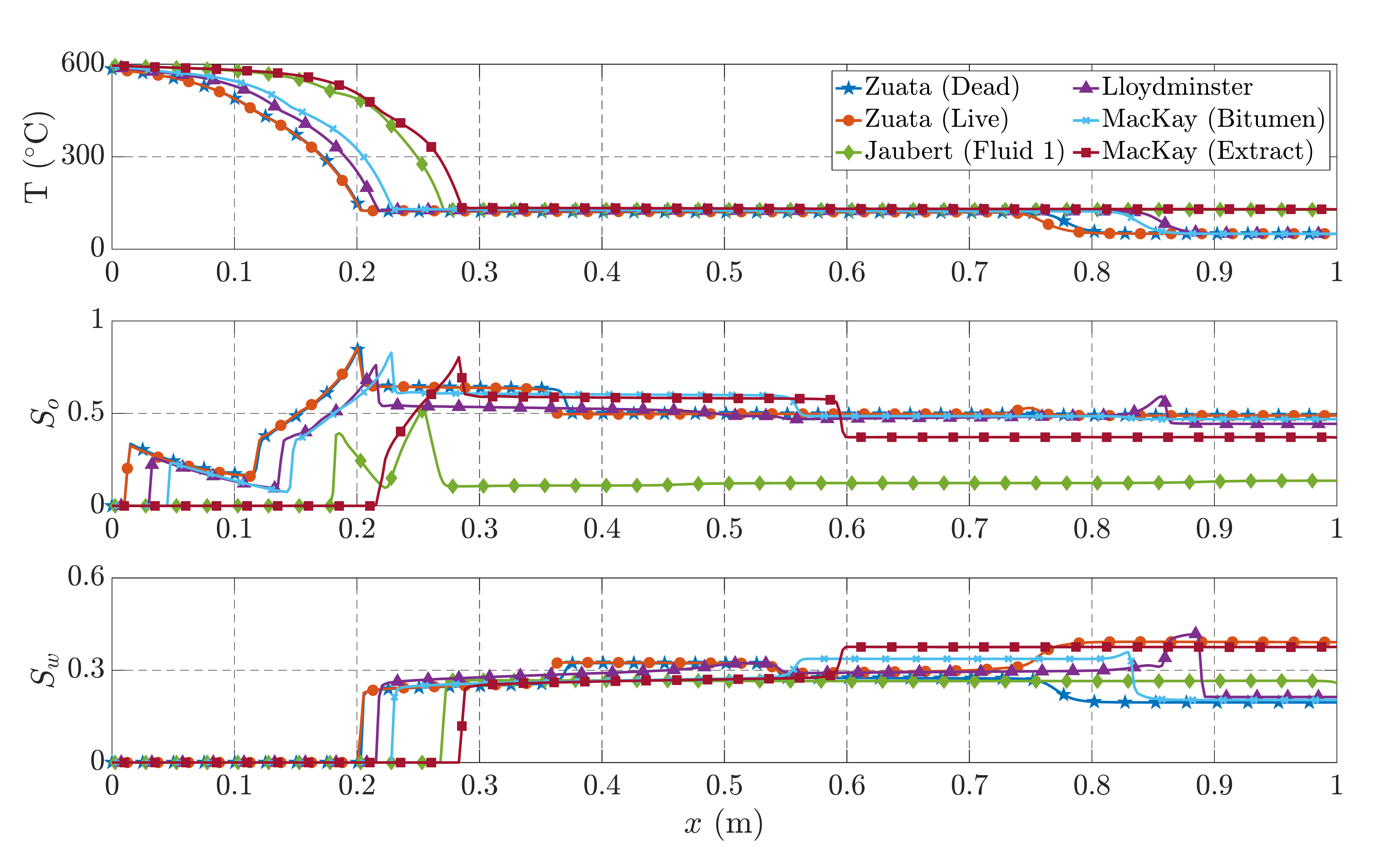}
    \caption{Temperature (top), oil saturation (center) and water saturation (bottom) profiles after 80 minutes of hot nitrogen injection, for all oil samples using the detailed, 52 hydrocarbon components.}
    \label{fig::nitrogenInj}
\end{figure*}

\begin{figure}[b!]
    \centering
    \includegraphics[width=.48\textwidth]{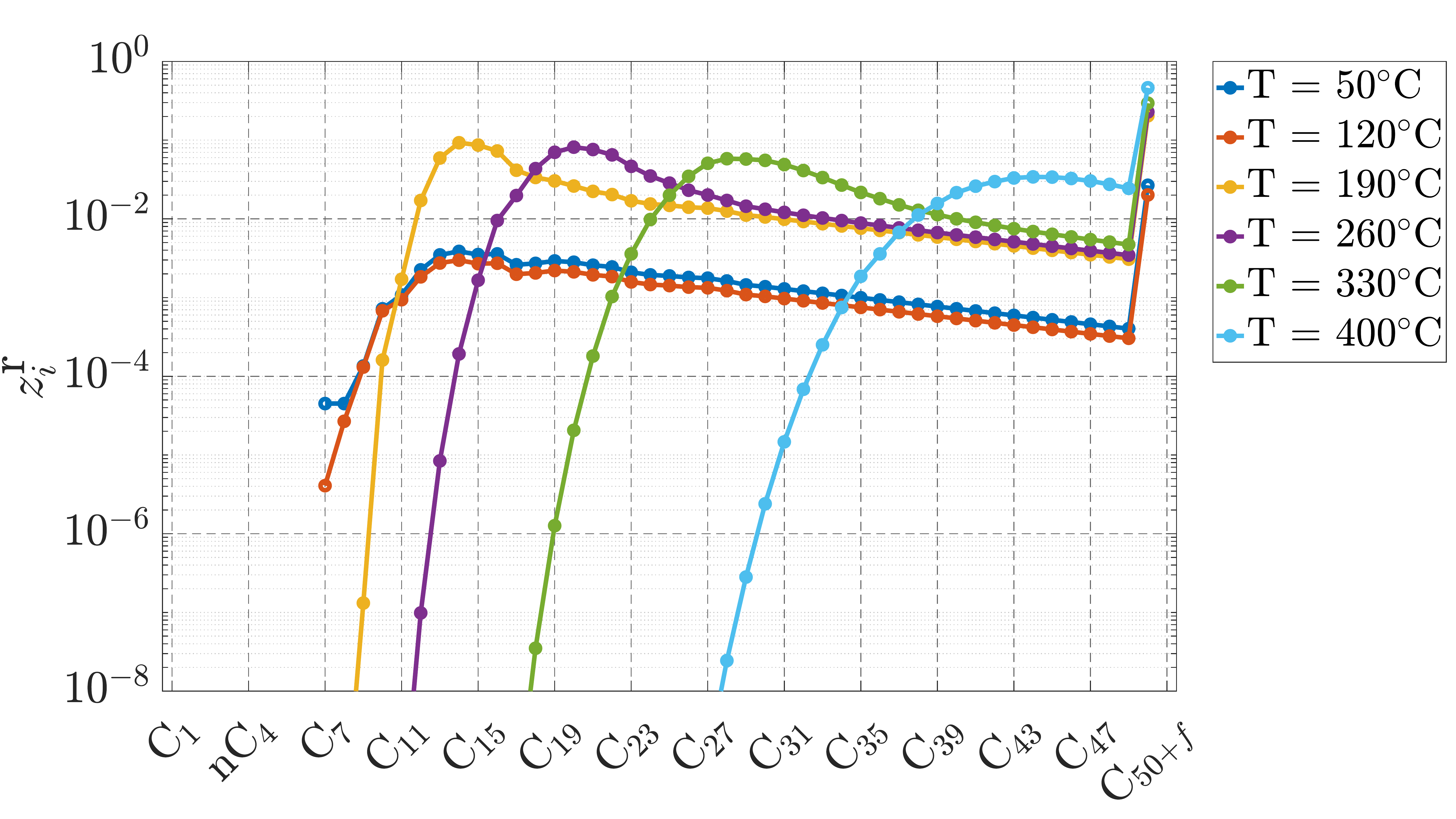}
    \caption{Reference hydrocarbon compositions for the Dead Zuata oil, nitrogen injection cases.}
    \label{fig::tempSelection}
\end{figure}

Figure~\ref{fig::nitrogenInj} shows the temperature, oil saturation and water saturation profiles after 80 minutes of hot nitrogen injection. We see several fronts moving downstream. The leading temperature front is the steam front, visible around 0.8m, followed by the mostly constant temperature steam plateau. Then the water phase gets evaporated, leading to an increased temperature from latent heat release and heat capacity effects, around 0.2-0.3m. Multiple displacement fronts can be seen on the oil saturation profile. The steam front displaces some light-medium components, with the Lloydminster oil showing the largest bank, and then the thermal bank made up from the components vaporized around the water evaporation front follows. Finally, all oils are fully evaporated by 600$^\circ$C, so we have trailing evaporation shocks next to the inlet. The figure further illustrates that thermal processes are sequential by nature. 
Figure~\ref{fig::tempSelection} shows the reference compositions using the uniform set for the Dead Zuata sample, spanning the composition space well and containing different information for each temperature.

\begin{figure*}[htb!]
    \centering
    \includegraphics[width=.95\textwidth]{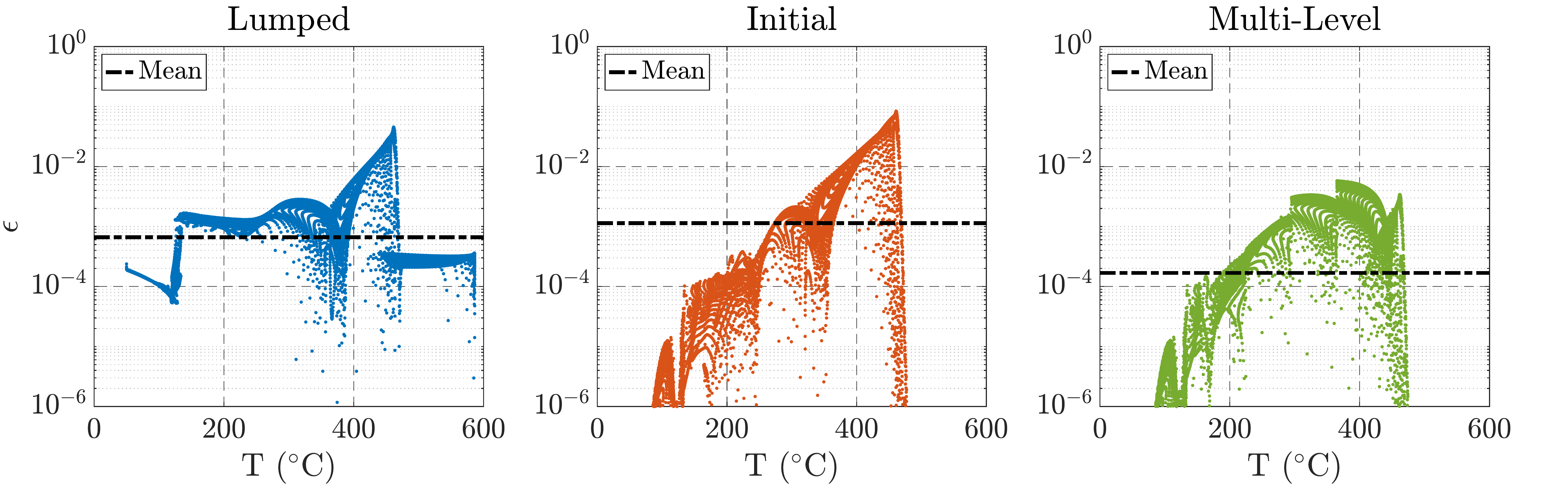}
    \includegraphics[width=.95\textwidth]{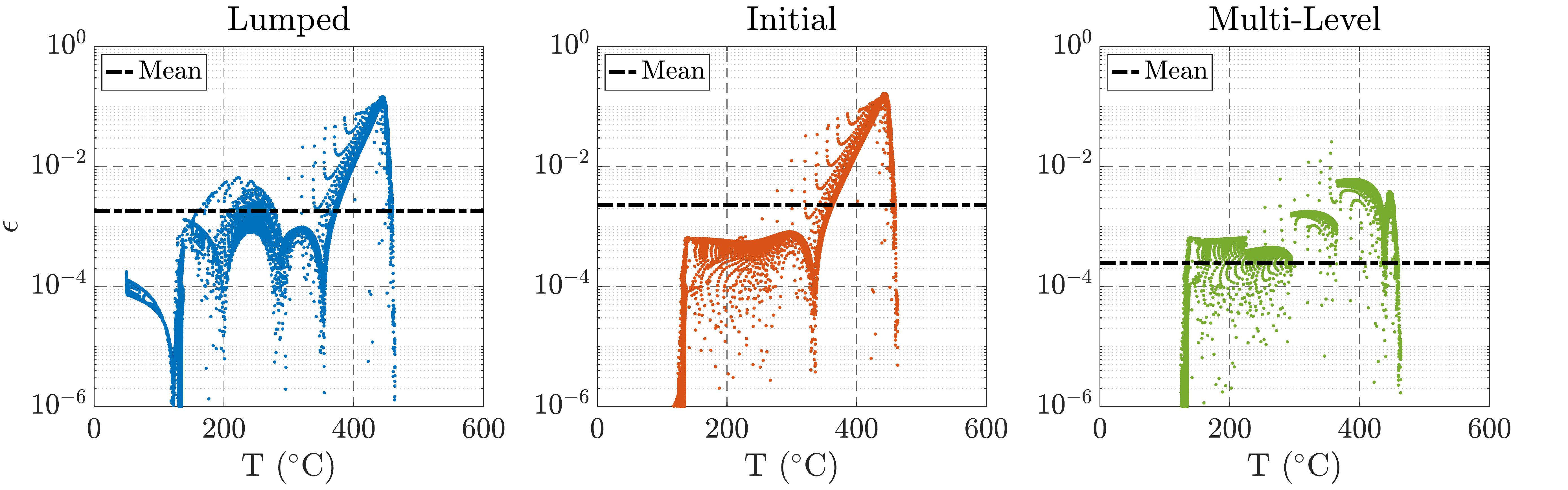}
    \includegraphics[width=.95\textwidth]{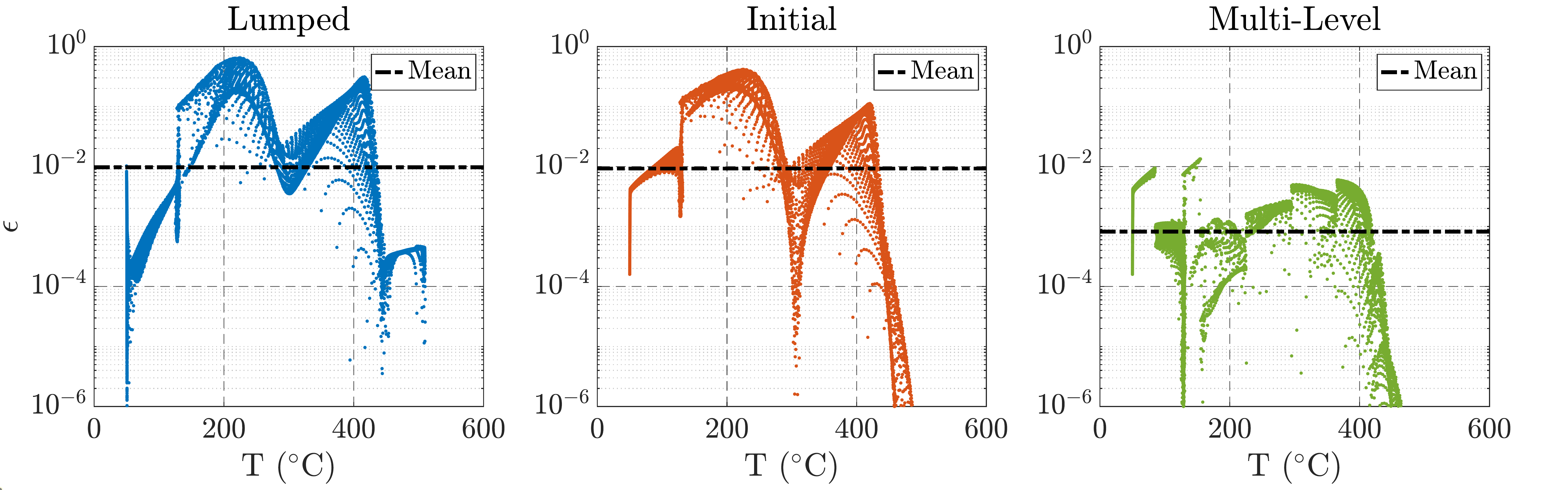}
    \caption{1-norm error on the liquid mole fraction (L), vapor mole fraction (V) and vapor compressibility factor (Z$_v$) as a function of temperature for the lumped composition (blue), delumping with the initial oil composition (orange), delumping using our multi-level strategy (green). Dead Zuata oil sample (top), MacKay Extract (center), Jaubert Fluid 1 (bottom).}
    \label{fig::scatter}
\end{figure*}

Figure~\ref{fig::scatter} shows the error on the liquid molar fraction using the lumped composition and our delumping methods (initial and multi-level) for three oils. We plot the error as a function of temperature, with each point a test sample. For the Dead Zuata oil, which is a typical extra-heavy oil profile, we observe that the error grows with temperature for the lumped flash, because the information contained in the heavy pseudo-component ($\textrm{C}_{17-49}$) is lost at high temperature. Using the initial composition
to delump reduced the error at low temperature, but above 200$^\circ$C its performance worsens, to the point of being worse than the lumped results. Our multi-level solution is able to lower the error across all temperatures. The lower temperature error benefits from the initial composition (corresponding to the initial temperature), but as temperature increases it switches to other reference compositions and is able to maintain the error below 1\%. We also note that as expected, the error shows discontinuities when we switch from one reference temperature to another.

One strength of our method is that it is amenable to including more information about the physics, both a priori and after we observe the results using a uniform set of temperatures. Figure~\ref{fig::scatter} (middle) shows the error as a function of temperature for the MacKay Extract oil. We see that unfortunately, our selection of temperatures leaves a few larger errors around 360$^\circ$C. We added 360$^\circ$C to the reference set, and doing so reduced the maximum error by a factor of 2.1 on top of the default set results. In Figure~\ref{fig::scatter} (bottom), we plot the results for the Jaubert Fluid 1 oil. In that case, the maximum errors for the multi-level method are seen at low temperature. The reason for that is the presence of light components in the initial oil, which will introduce artefacts due to those components being stripped away in the very first few timesteps of the simulation. To fix that problem, we introduced 51$^\circ$C in the set of temperatures, and were able to further reduce the average error by a factor of 2.2. 

The default, uniform set can always be used as a first guess and leads to good results on average, but there are multiple ways to increase the accuracy of the method by adding more knowledge about the physics.
\begin{figure*}[t!]
    \centering
    \includegraphics[width=\textwidth]{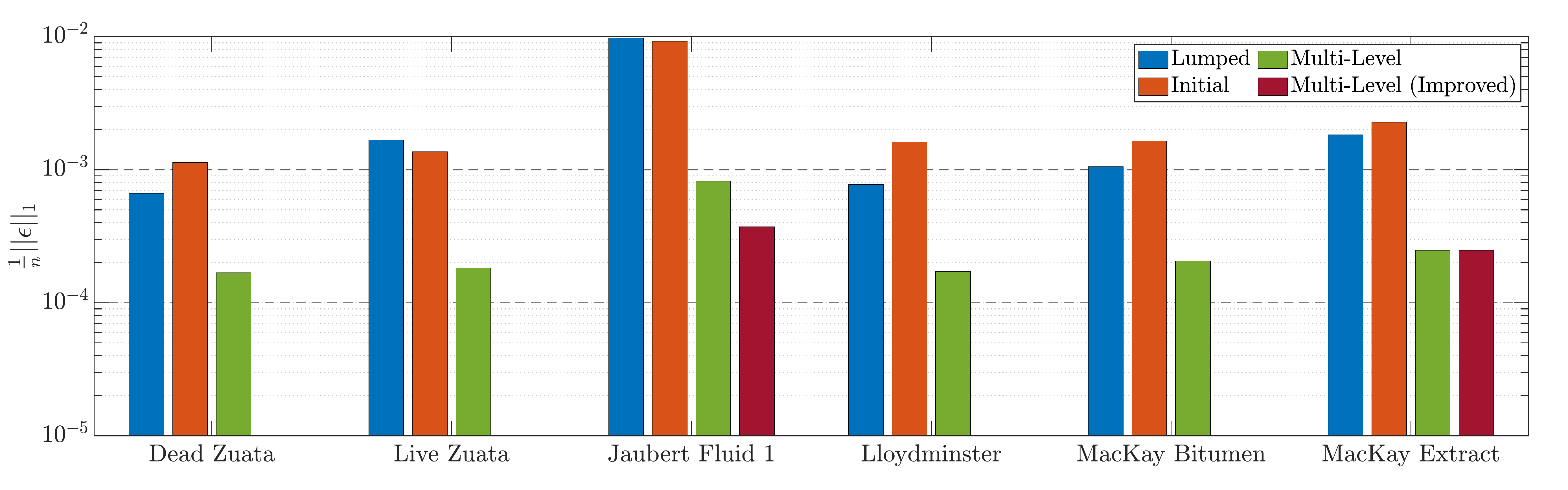}
    \includegraphics[width=\textwidth]{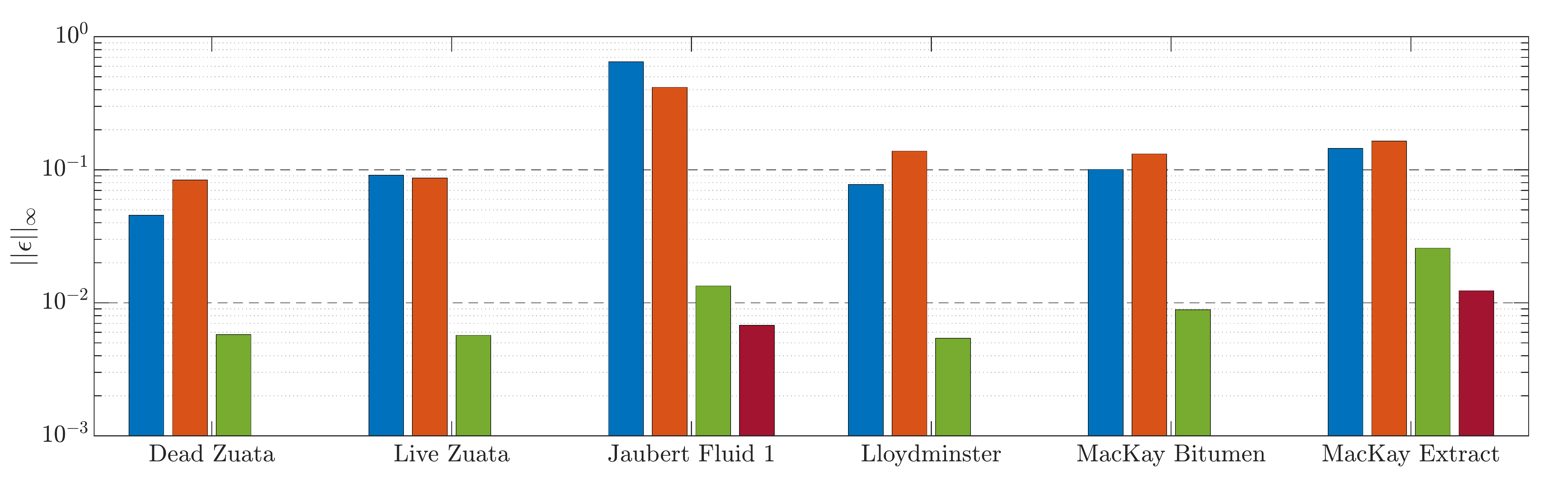}
    \caption{Average 1-norm error (top) and maximum error (bottom) on the liquid mole fraction (L) for all oil samples under hot nitrogen injection, plotted on a logarithmic scale. The improved results are only shown for samples to which we applied a modification.}
    \label{fig::errorL}
\end{figure*}
Figure~\ref{fig::errorL} shows the average (top) and maximum (bottom) error on the liquid mole fraction for all oils. A full breakdown of the default temperature set results is given in Table~\ref{tab::results}. The absolute error is kept below 1\% for all cases but the MacKay Extract oil and Jaubert Fluid 1 oils. If we implement the slight modifications mentioned above (denoted Improved), the average error on the liquid mole fractions across all oil samples is at least four times smaller than using lumped results, and on average seven times. The maximum error is reduced by at least six times, and on average 17 times.

\subsection{Test Case 2: In-situ Combustion}

We now consider our target application, in-situ combustion. We inject air at 400$^\circ$C, triggering the oxidation reactions. We use a three-reaction scheme from \citet{Dechelette06}:
\begin{subequations}
\begin{align}
 \textrm{Oil} + \alpha_1 \textrm{O$_2$} & \rightarrow \beta_1 \textrm{Coke$_1$},\\
 \textrm{Coke$_1$} + \alpha_2 \textrm{O$_2$} & \rightarrow \beta_2 \textrm{Coke$_2$} + \gamma_2 \textrm{CO$_2$} + \delta_2  \textrm{H$_2$O}, \\
 \textrm{Coke$_2$} + \alpha_3 \textrm{O$_2$} & \rightarrow \gamma_3 \textrm{CO$_2$} + \delta_3  \textrm{H$_2$O},
\end{align}
\end{subequations}
where $\alpha_i$, $\beta_i$, $\gamma_i$ and $\delta_i$ are coefficients ensuring the mass balance. The heavy fraction of the oil $(\textrm{C}_{50+}$, denoted Oil in the reaction scheme$)$ is converted to solid fuel (Coke$_1$) at medium temperatures, and that fuel is burned at high temperatures. In turn, it produces an oxidized solid fuel (Coke$_2$), which will burn at even higher temperatures. The reaction parameters are given in Table~\ref{tab::ISCparams}. 

\begin{table}[b!]
    \centering
    \caption{Reaction parameters for the \citet{Dechelette06} model. Activation energy (E$_\textrm{a}$), frequency factor (A) and enthalpy of reaction $(h_r)$.}
    {
    \begin{tabular}{lrrr}
    \toprule
    Reac. & E$_\textrm{a}$ (J/mol)  & A (min$^{-1}$.kPa$^{-1}$) & $h_r$ (GJ) \\
    \midrule
    1 & 25,000 & 1 & 1.60 \\
    2 & 67,000 & 250 & 12.80 \\
    3 & 87,000 & 220 & 4.85 \\
    \bottomrule
    \end{tabular}
    }
    \label{tab::ISCparams}
\end{table}

\begin{figure}[b!]
    \centering
    \includegraphics[width=.48\textwidth]{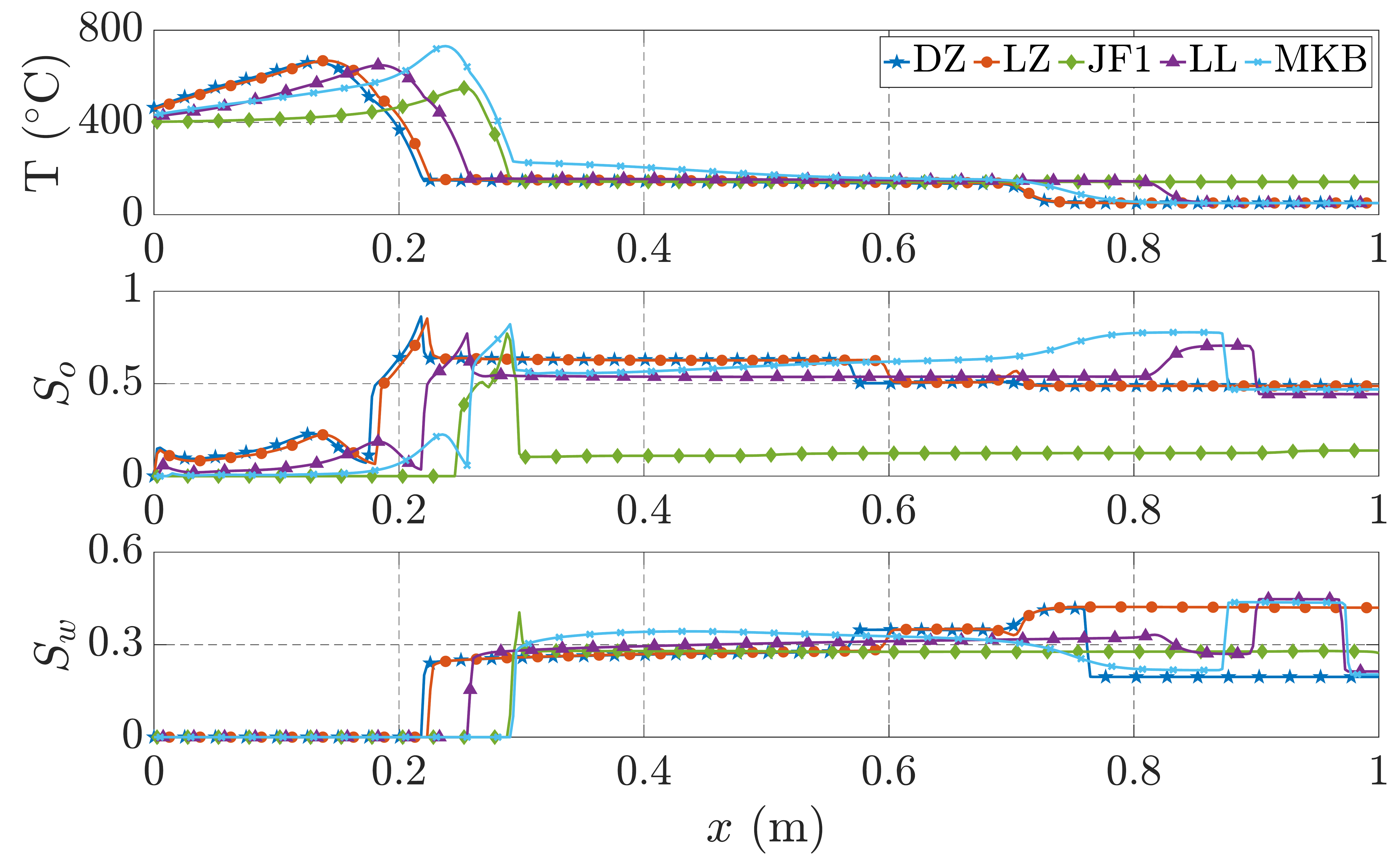}
    \caption{Temperature (top), oil saturation (center) and water saturation (bottom) profiles after 80 minutes of air injection, for all oil samples using the detailed, 52 hydrocarbon components for Dead Zuata (DZ), Live Zuata (LZ), Jaubert Fluid 1 (JF1), Lloydminster (LL) and MacKay Bitumen (MKB).}
    \label{fig::airInj}
\end{figure}

\begin{figure*}[t!]
    \centering
    \includegraphics[width=\textwidth]{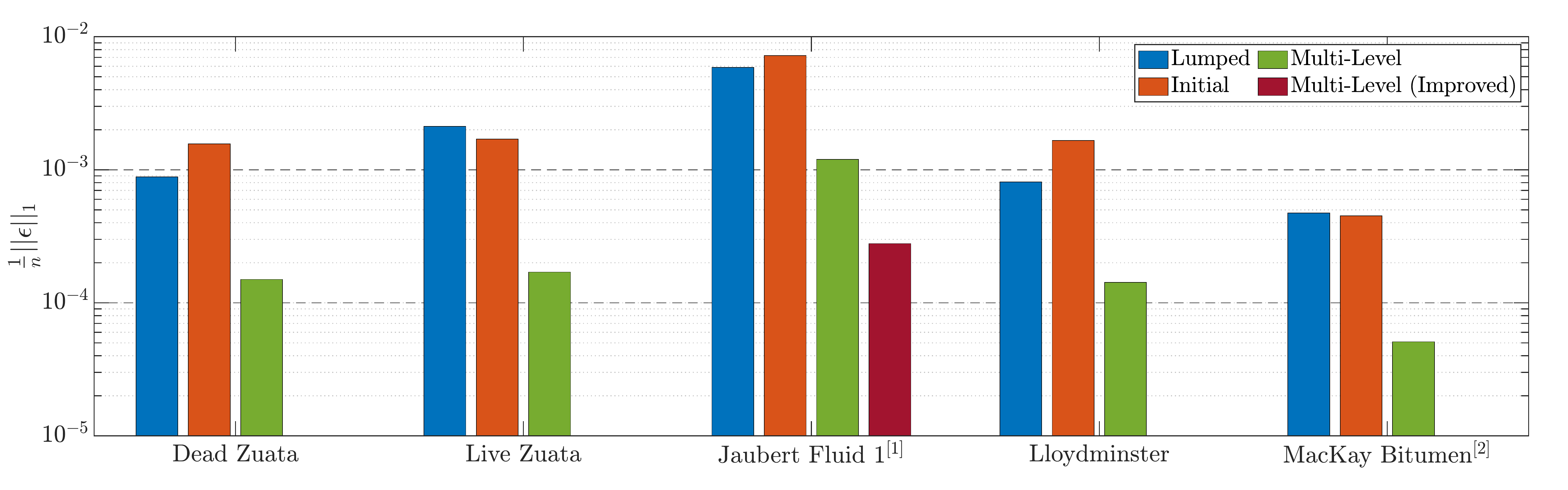}
    \includegraphics[width=\textwidth]{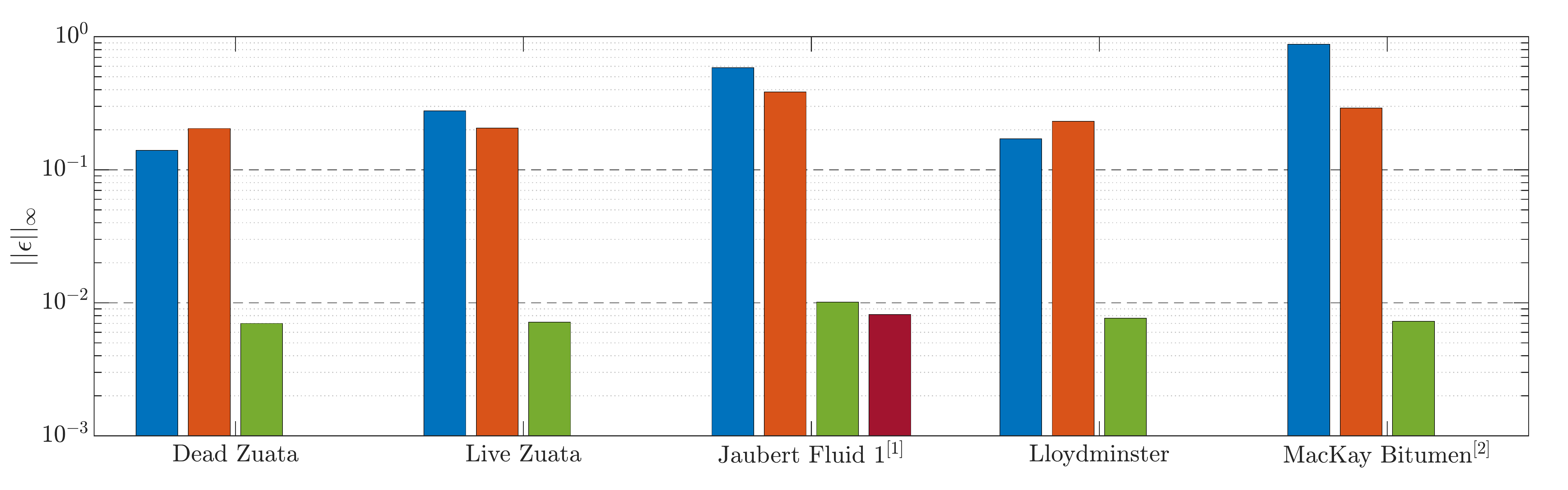}
    \caption{Average 1-norm error (top) and maximum error (bottom) on the liquid mole fraction (L) for all oil samples under air injection (in-situ combustion), plotted on a logarithmic scale. The improved results are only shown for samples to which we applied a modification. $^{[1]}$ results obtained over 151 timesteps. $^{[2]}$ results obtained over 81 timesteps.}
    \label{fig::errorLISC}
\end{figure*}

Following the conclusions of \citet{Kovscek13}, we burn up to 8\% of the total mass. Every other simulation property is the same as in the previous section. Figure~\ref{fig::airInj} shows the simulation results after 80 minutes of air injection. Note that we do not consider MacKay Extract for this test case, because the plus fraction is far less than 8\% by mass, preventing us from a fair comparison with the other oils. The main difference between air and nitrogen injection is obviously the presence of chemical reactions. We see that the maximum temperature is higher than the injection temperature of 400$^\circ$C, due to the enthalpy of reaction. The combustion front represents both the maximum reaction rate and maximum temperature. We see that the oil is displaced efficiently, with a large bank forming downstream of the combustion front. The presence of reactions leads to a much faster temperature increase. For our delumping method, it means that the components have less time to be vaporized, and we should use higher temperatures to better span the compositional space. We again use six temperatures, evenly distributed between 50$^\circ$C and 450$^\circ$C. A full breakdown of the results are given in Table~\ref{tab::results}. In Figure~\ref{fig::errorLISC} we plot the liquid molar fraction errors for all oils considered. Once again, we add a low temperature (51$^\circ$C) to improve the results for the Jaubert Fluid 1 oil, and lowered the average error by another 4.3x factor. The sharper temperature increase and the presence of reactions (which lowers the mole fraction of C$_{50+}$ and changes the overall profile of the compositions) are adding more challenges for the lumped simulations. Our method performs even better here than for the nitrogen injection case; the average error is reduced at least five times and the maximum error at least twenty times.

\subsection{Sensitivity Study}

\begin{figure*}[t!]
    \centering
    \includegraphics[width=\textwidth]{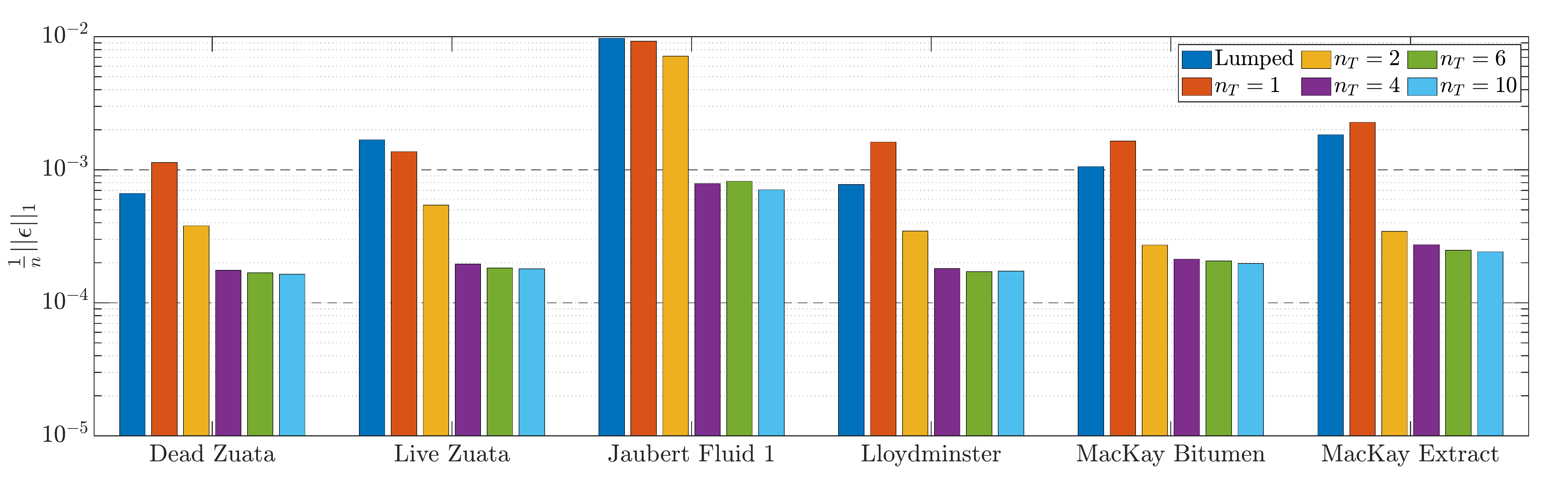}
    \includegraphics[width=\textwidth]{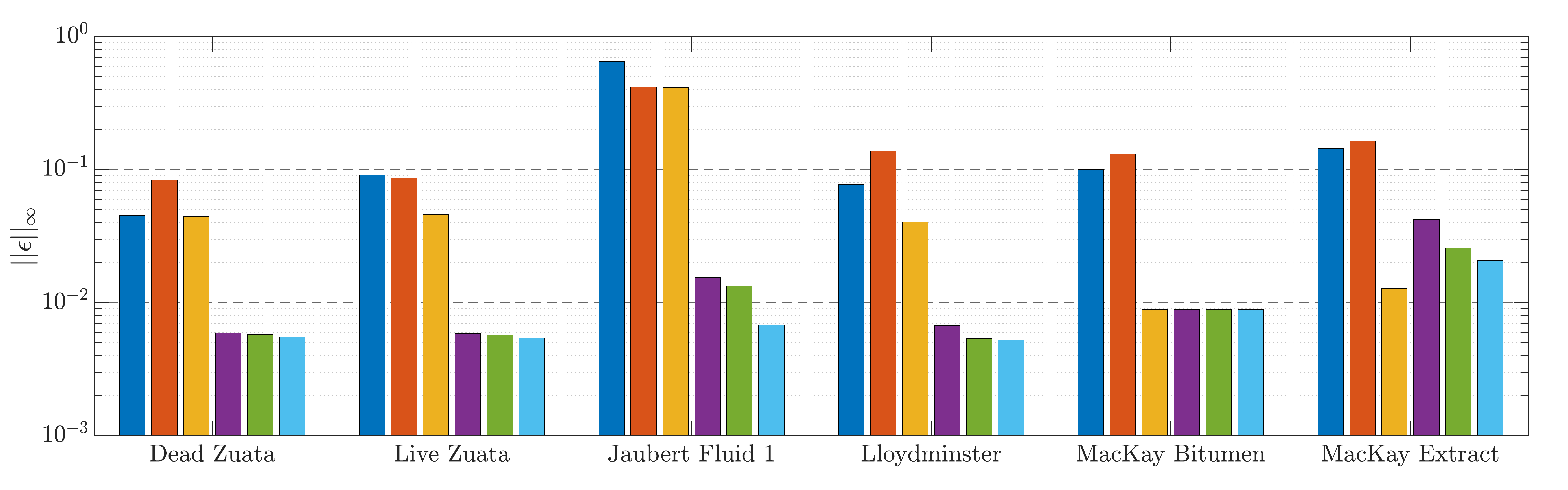}
    \caption{Average 1-norm error (top) and maximum error (bottom) on the liquid mole fraction (L) for all oil samples under nitrogen injection, plotted on a logarithmic scale using two (yellow), four (purple), six (green) and ten (light blue) reference temperatures.}
    \label{fig::temps}
\end{figure*}

We conducted a sensitivity study to gauge the impact of using different number of reference temperatures. Even though the storage requirements are limited (one vector per temperature), a large number of temperatures ($n_T$) is unlikely to yield much improvement. The missing information we still encounter by using our proxy method cannot be fully recovered by simply adding more temperatures, but using too few can lead to poor performance on the more challenging oils. Figure~\ref{fig::temps} shows the results for all oils using the nitrogen injection test case, using two, four, six and ten reference temperatures. The average error converges the fastest, showing virtually no improvement above four temperatures. Using six or ten temperatures can improve the performance on the maximum norm for more challenging oils, like Jaubert Fluid 1 and MacKay Extract. For those reasons, we selected six temperatures as our base case. As we previously noted, addressing specific issues with the maximum error can be done manually by looking at the uniform set errors and adding specific temperatures. Table~\ref{tab::temps} summarizes the sensitivity study results for both test cases.

Finally, we note that the run time for the method is independent of the number of temperatures used in the set. It only impacts which temperature is selected as reference, but the number of flashes run is constant.

\begin{table}[hb!]
    \centering
    \caption{Liquid molar fraction error ($\epsilon$) for different number of uniformly sampled reference temperatures. Results are averaged across all six oil samples, from the nitrogen (Nit.) injection case (top) and air injection case (bottom). The relative improvement (R.I.) is based on the 1-temperature results (Initial).}
    {
    \begin{tabular}{llrrrr}
    \toprule
       & $n_T$ & Mean $\epsilon$ & R.I. & Max $\epsilon$ & R.I.  \\
    \midrule
    \multirow{5}{*}{Nitrogen}  &  1 & 0.00288 & -- & 0.17022 & -- \\
       & 2 & 0.00151 & 66.1$\%$ & 0.09481 & 58.3$\%$ \\
       & 4 & 0.00030 & 87.6$\%$ & 0.01421 & 90.8$\%$\\
       & 6 & 0.00030 & 88.1$\%$ & 0.01083 & 92.8$\%$ \\
       & 10 & 0.00028 & 88.6$\%$ & 0.00878 & 93.7$\%$\\
        \midrule
       \multirow{5}{*}{Air} & 1 & 0.00252 & -- & 0.26335 & -- \\
       & 2 & 0.00159 & 68.7$\%$ & 0.08782 & 75.2$\%$ \\
       & 4 & 0.00039 & 87.7$\%$ & 0.00961 & 96.5$\%$\\
       & 6 & 0.00033 & 89.0$\%$ & 0.00750 & 97.0$\%$ \\
       & 10 & 0.00033 & 89.1$\%$ & 0.00730 & 97.1$\%$\\
    \bottomrule
    \end{tabular}
    }
    \label{tab::temps}
\end{table}

\subsection{Optimization}
\label{sec::optim}

\begin{figure*}[t!]
    \centering
    \includegraphics[width=\textwidth]{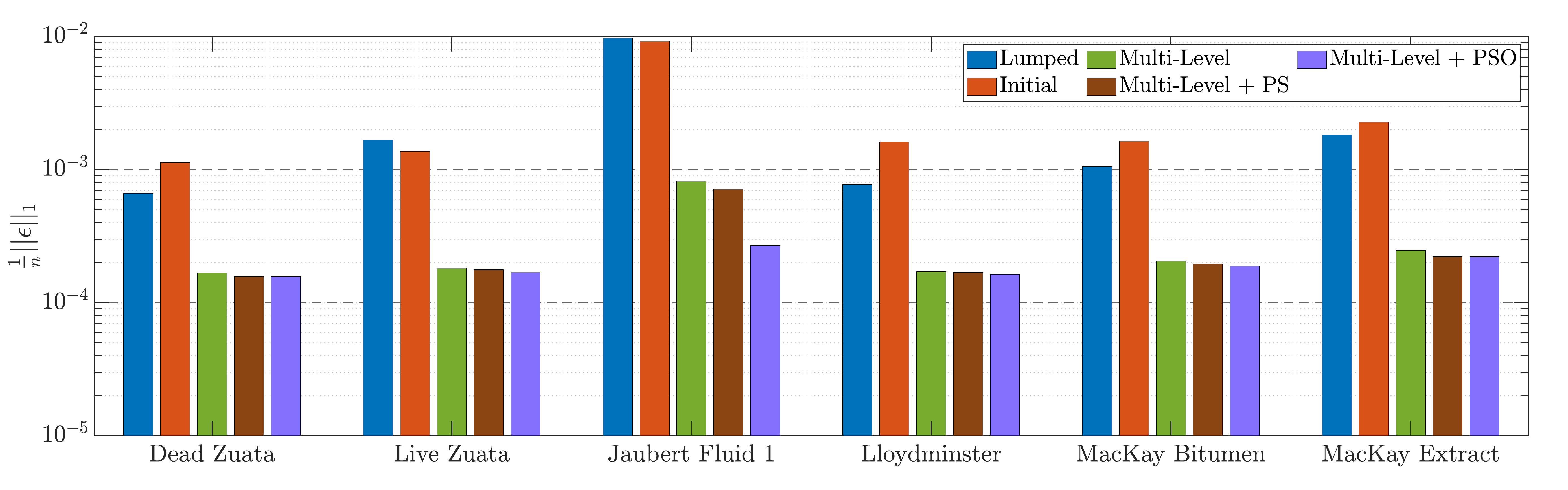}
    \includegraphics[width=\textwidth]{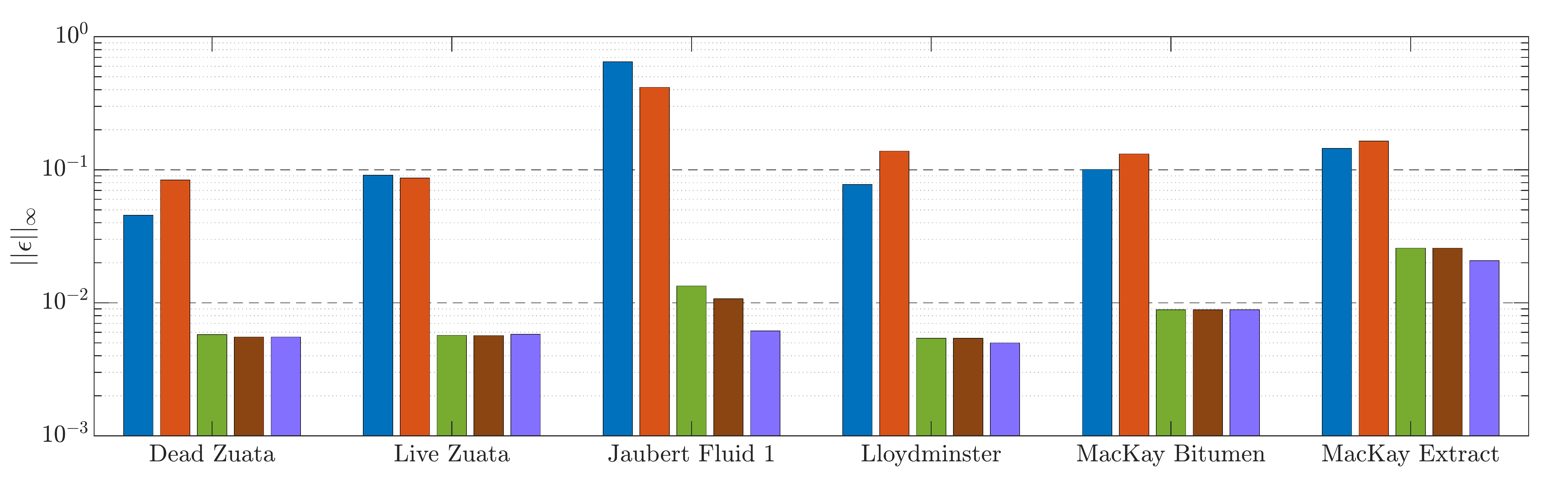}
    \caption{Average 1-norm error (top) and maximum error (bottom) on the liquid mole fraction (L) for all oil samples under nitrogen injection, plotted on a logarithmic scale. We compare the uniform set of six temperatures (green) with results from Pattern Search (PS, brown) and Particle Swarm Optimization (PSO, lavender).}
    \label{fig::errorLoptim}
\end{figure*}

We then considered improving the results via optimization methods. The error space is non convex and we have no access to analytical gradients, so we decided to use two gradient-free methods. Pattern Search (PS, \citep{Hooke60,Davidon91}) is a heuristic method that scans points at a fixed step size around the current location, and moves along to the best new trial. If no new trial is better, it reduces the step size and repeats the process. From its formulation, it can find a local minimum for a low computational cost, but will not be able to move far in the error space to find the global minimum. Particle Swarm Optimization (PSO, \citep{Kennedy95}), on the other hand, has no guarantee to find even a local minimum (at least in its vanilla formulation), but if we can afford a large number of particles, it can find the global minimum. In this study, the cost function evaluation involves 80,400 full flashes and the dimension of the search space is between two and eight, making PSO an extremely expensive method. To reduce the number of variables in the optimization problem, we keep the minimum and maximum temperatures of the set fixed. For PSO in particular, which suffers greatly from the curse of dimensionality, it will lead to more meaningful results for a fixed cost. We always optimize using the average 1-norm error on the liquid mole fraction as the cost function, only subjecting the variables to be within the minimum and maximum temperatures
\begin{subequations}
\begin{align}
        & \min\limits_{\textrm{T}} \dfrac{1}{n}||\epsilon||_1 \\
    \text{s.t.} \quad & \textrm{T}_{\textrm{min}} \leq \textrm{T}_i \leq \textrm{T}_{\textrm{min}}, \quad \forall i = 1,\dots,n_T-2
\end{align}
\end{subequations}

We ran PS on all oil samples for the nitrogen injection, with an initial temperature step of 20$^\circ$C, a minimum step of 1$^\circ$C and using the uniform set as the initial guess. Jaubert Fluid 1 shows the best improvement with 12.6\%, and the average is around 6.5\%. As expected, PS is not capable of finding a better convex region than the one containing the initial guess. The nature of the delumping method makes the uniform set a good option regardless of the oil sample and test case, and our PS results suggests that it sits in a mostly flat region of the search space.

The global exploration ability of PSO can potentially find better results by jumping to a better convex region, at the expense of a much larger cost. We used 100 particles for our runs, and a maximum of 15 iterations. We added a PS run with an initial temperature step of 2$^\circ$C on the final solution, to make sure we reached a local minimum. For Jaubert Fluid 1, we already noted in Section~\ref{subsec::nit} that the presence of light components in the initial oil was throwing off the method at low temperatures. PSO was able to capture that, leading to a 67.2\% improvement (or 3x) compared to the uniform set. Our manual improvement got us a 2.2x improvement, so in that case PSO is able to provide a better solution. All other cases show comparable performance to PS, for a much larger computational cost (around 10-20x depending on the cases). Table~\ref{tab::optimResults} summarizes the improvement from our optimization runs, and Figure~\ref{fig::errorLoptim} plots them.

\begin{table}[ht!]
    \centering
    \caption{Liquid molar fraction error ($\epsilon$) using Pattern Search (PS) and Particle Swarm Optimization (PSO), for the nitrogen injection case. The relative improvement (R.I.) is based on the six temperatures uniform set.}
    {
    \begin{tabular}{llrr}
    \toprule
         &  & Mean Error & Max Error  \\
        Method & Oil & Improvement & Improvement  \\
    \midrule
         & DZ & 6.5\% &  4.1\% \\
        & LZ &  2.9\% &  0.5\% \\
        Pattern & JF1 &  12.6\% & 20.0\% \\
        Search & LL &  1.7\% & 0.1\% \\
        & MKB & 4.9\% & 0.0\% \\
        & MKE &  10.4\% & 0.0\% \\
        \midrule
        & DZ & 6.5\% & 4.1\%\\
        & LZ & 7.0\% & 1.8\% \\
        Particle Swarm & JF1 & 67.2\% & 54.1\% \\
        Optimization & LL & 4.9\% & 7.8\%\\
        & MKB & 8.2\% & 0.0\% \\
        & MKE & 10.7\% & 19.5\% \\
    \bottomrule
    \end{tabular}
    }
    \label{tab::optimResults}
\end{table}

\section{Conclusion \& Discussion}

We presented a new delumping method for thermal recovery processes. We tested the method on two different thermal recovery cases, for six oil samples. At low pressure, we lower the average error made on the phase molar fractions by an order of magnitude (4-12x reduction, average of 7x). The maximum error is further reduced, by 6-48x and an average of 17x. We observe even better results on the more challenging in-situ combustion cases, with average reductions of 7.6x for the 1-norm error and 52x for the maximum error.

The method is amenable to adding more knowledge about the physics and/or to be paired with gradient-free optimization methods. The local optimizer Pattern Search yields an average 6.5\% improvement over our base case results. The global Particle Swarm Optimization algorithm is expensive to run but has, in at least one case, lead to a large improvement of 67\%.

This work provides a new direction for the simulation of complex, heavy crude oils displacement processes under thermal recovery methods. The large number of (pseudo-)components needed to simulate those cases leads to a prohibitively expensive computational cost. Our method could bridge the gap between the accuracy of the flash process and the number of components retained in the global, non-linear Newton solver.

As for future research avenues, we would also like to test the method at field pressure to challenge the low pressure hypothesis.
Running two- and three-dimensional cases would allow us to investigate if a 1D detailed model is enough to guide higher dimension cases \citep{Rannou13}.
Integrating the algorithm into a robust, molar variables based reservoir simulator would allow for further testing.

\section*{Acknowledgements}
\label{sec::acknow}

The authors wish to thank Prof. Anthony R. Kovscek (Stanford), Prof. Hamdi A. Tchelepi (Stanford), Prof. Dan V. Nichita (Universit\'e de Pau et des Pays de l'Adour) and Dr. Jacques Franc (Stanford) for numerous suggestions and fruitful discussions, and Ecopetrol for financial support.

\bibliographystyle{elsarticle-num-names}

\bibliography{bib_Delumping}

\appendix

\section{Table of Results}

\begin{table*}[b!]
    \centering
    \caption{Average 1-norm (Mean) and infinity-norm (Max) errors on the liquid mole fraction $(\epsilon)$, for the nitrogen injection cases and air injection cases and all oil samples. Ratio denotes the lumped error divided by the multi-level error (using six temperatures).}
    {
    \begin{tabular}{lllccrccr}
    \toprule
     & & & \multicolumn{3}{c}{Nitrogen Injection} & \multicolumn{3}{c}{Air Injection} \\
    \cmidrule{4-9}
    Oil & Estimator & Quantity & Lumped & Multi-Level & Ratio & Lumped & Multi-Level & Ratio \\
    \midrule
    \multirow{4}{*}{Dead Zuata} & \multirow{2}{*}{Mean} & L & 6.6436e$^{-4}$ & 1.6841e$^{-4}$ & 4.0 & 8.8387e$^{-4}$ & 1.4994e$^{-4}$ & 5.9 \\
     & & $Z_v$ & 8.6751e$^{-4}$ & 2.4819e$^{-5}$ & 35.0 & 1.1831e$^{-3}$ & 6.6945e$^{-5}$ & 17.7  \\
     \cmidrule{2-9}
     & \multirow{2}{*}{Max} & L & 4.5519e$^{-2}$ & 5.7726e$^{-3}$ & 7.9 & 1.4001e$^{-1}$ & 6.9983e$^{-3}$ & 20.0 \\
     & & $Z_v$ & 1.0210e$^{-2}$ & 1.3086e$^{-3}$ & 7.8 & 4.7470e$^{-1}$ & 3.7230e$^{-3}$ & 12.8  \\
    \midrule
    
    \multirow{4}{*}{Live Zuata} & \multirow{2}{*}{Mean} & L & 1.6806e$^{-3}$ & 1.8289e$^{-4}$ & 9.2 & 2.1160e$^{-3}$ & 1.7025e$^{-4}$ & 12.4  \\
     & & $Z_v$ & 3.9829e$^{-3}$ & 1.7225e$^{-3}$ & 23.1 & 4.6772e$^{-3}$ & 2.1932e$^{-4}$ & 21.3  \\
     \cmidrule{2-9}
     & \multirow{2}{*}{Max} & L & 9.1150e$^{-2}$ & 5.7045e$^{-3}$ & 16.0 & 2.7737e$^{-1}$ & 7.1622e$^{-3}$ & 38.7  \\
     & & $Z_v$ & 8.5913e$^{-2}$ & 2.6001e$^{-3}$ & 33.0 & 1.2351e$^{-1}$ & 6.1811e$^{-3}$ & 20.0  \\
    \midrule
    
    \multirow{4}{*}{Jaubert Fluid 1} & \multirow{2}{*}{Mean} & L & 9.7412e$^{-3}$ & 8.1988e$^{-4}$ & 11.9 & 6.0693e$^{-3}$ & 1.3012e$^{-3}$ & 4.7 \\
     & & $Z_v$ & 3.9829e$^{-3}$ & 1.7225e$^{-4}$ & 9.0 & 2.0790e$^{-2}$ & 3.5393e$^{-3}$ & 5.9 \\
     \cmidrule{2-9}
     & \multirow{2}{*}{Max} & L & 6.4768e$^{-1}$ & 1.3395e$^{-2}$ & 48.4 & 5.8485e$^{-1}$ & 1.0132e$^{-2}$ & 57.7 \\
     & & $Z_v$ & 1.3997e$^{-1}$ & 2.3600e$^{-2}$ & 5.9 & 1.4667e$^{-1}$ & 2.5015e$^{-2}$ & 5.9 \\
    \midrule
    
    \multirow{4}{*}{Lloydminster} & \multirow{2}{*}{Mean} & L & 7.7658e$^{-4}$ & 1.7172e$^{-4}$ & 4.5 & 8.0964e$^{-4}$ & 1.4224e$^{-4}$ & 5.7 \\
     & & $Z_v$ & 9.7628e$^{-4}$ & 5.0994e$^{-5}$ & 5.7 & 1.1374e$^{-3}$ & 7.7672e$^{-5}$ & 14.6 \\
     \cmidrule{2-9}
     & \multirow{2}{*}{Max} & L & 7.7436e$^{-2}$ & 5.4227e$^{-3}$ & 14.3 & 1.7090e$^{-1}$ & 7.6602e$^{-3}$ & 22.3 \\
     & & $Z_v$ & 1.7545e$^{-2}$ & 1.3886e$^{-3}$ & 19.1 & 7.0511e$^{-2}$ & 3.2109e$^{-3}$ & 22.0 \\
     
    \midrule
    \multirow{4}{*}{MacKay Bitumen} & \multirow{2}{*}{Mean} & L & 1.0571e$^{-3}$ & 2.0614e$^{-4}$ & 5.1 & 4.7285e$^{-4}$ & 5.1007e$^{-5}$ & 9.3 \\
     & & $Z_v$ & 1.4373e$^{-3}$ & 3.6698e$^{-5}$ & 39.2 & 1.3628e$^{-3}$ & 3.2072e$^{-5}$ & 42.5 \\
     \cmidrule{2-9}
     & \multirow{2}{*}{Max} & L & 1.0068e$^{-1}$ & 8.8686e$^{-3}$ & 11.4 & 8.7758e$^{-1}$ & 7.2650e$^{-3}$ & 120.8 \\
     & & $Z_v$ & 2.6089e$^{-2}$ & 1.1518e$^{-3}$ & 22.7 & 4.6377e$^{-1}$ & 5.0607e$^{-3}$ & 91.6 \\
    \midrule
         
    \multirow{4}{*}{MacKay Extract} & \multirow{2}{*}{Mean} & L & 1.8344e$^{-3}$ & 2.4825e$^{-4}$ & 7.4 & -- & -- & -- \\
     & & $Z_v$ & 2.5257e$^{-3}$ & 1.0319e$^{-4}$ & 24.5 & -- & -- & -- \\
     \cmidrule{2-9}
     & \multirow{2}{*}{Max} & L & 1.4499e$^{-1}$ & 2.5790e$^{-2}$ & 5.6 & -- & -- & -- \\
     & & $Z_v$ & 8.6288e$^{-1}$ & 2.3124e$^{-3}$ & 373.2 & -- & -- & -- \\
     
    \bottomrule
    \end{tabular}
    }
    \label{tab::results}
\end{table*}

\end{document}